\documentclass[final,3p,times]{elsarticle}
\usepackage{amssymb}
\usepackage{amsmath}
\usepackage{multirow}
\usepackage{xcolor}
\usepackage{graphicx}
\usepackage{subcaption}
\newcommand{\myrev}[1]{\textcolor{black}{#1}}
\newcommand{\myrevtwo}[1]{\textcolor{black}{#1}}

\journal{ Internet of Things; Engineering Cyber Physical Human Systems}

\begin{document}

\begin{frontmatter}


%\title{Title\tnoteref{label1}}

\title{A Novel Federated Learning-Based IDS for \\ Enhancing UAVs Privacy and Security}


%% uncommented the following author names to submit arxiv -- Pinar
\author[address]{Ozlem Ceviz}
\

\author[address]{Pinar Sadioglu}

\author[address]{Sevil Sen \corref{mycorrespondingauthor}}
\ead{ssen@cs.hacettepe.edu.tr}
\address[address]{WISE Lab., Dept. of Computer Engineering, Hacettepe University, Ankara, Turkey}
\cortext[mycorrespondingauthor]{Corresponding author}

\author[label3]{Vassilios G. Vassilakis}
 \affiliation[label3]{organization={Department of Computer Science, University of York},
            city={York},
            postcode={YO10 5GH},
            country={United Kingdom}}
%\author{Ozlem Ceviz, Pinar Sadioglu, Sevil Sen, } %% Author name


%% Abstract
\begin{abstract}
 Unmanned aerial vehicles (UAVs) operating within Flying Ad-hoc Networks (FANETs) encounter security challenges due to the dynamic and distributed nature of these networks. Previous studies focused predominantly on centralized intrusion detection, assuming a central entity responsible for storing and analyzing data from all devices. However, these approaches face challenges including computation and storage costs, along with a single point of failure risk, threatening data privacy and availability. The widespread dispersion of data across interconnected devices underscores the need for decentralized approaches. This paper introduces the Federated Learning-based Intrusion Detection System (FL-IDS), addressing challenges encountered by centralized systems in FANETs. FL-IDS reduces computation and storage costs for both clients and the central server, which is crucial for resource-constrained UAVs. Operating in a decentralized manner, FL-IDS enables UAVs to collaboratively train a global intrusion detection model without sharing raw data, thus avoiding delay in decisions based on collected data, as is often the case with traditional methods. Experimental results demonstrate FL-IDS's competitive performance with Central IDS (C-IDS) while mitigating privacy concerns, with the Bias Towards Specific Clients (BTSC) method further enhancing FL-IDS performance even at lower attacker ratios. Comparative analysis with traditional intrusion detection methods, including Local IDS (L-IDS), sheds light on the strengths of FL-IDS. This study significantly contributes to UAV security by introducing a privacy-aware, decentralized intrusion detection approach tailored to UAV networks. Moreover, by introducing a realistic dataset for FANETs and federated learning, our approach differs from others lacking high dynamism and 3D node movements or accurate federated data federations.
\end{abstract}


\end{frontmatter}



\section{Introduction}
\label{sec1}
Unmanned aerial vehicles (UAVs) deployed in flying ad hoc networks (FANETs) \cite{bekmezci2013flying}  have attracted significant attention for their diverse applications, spanning surveillance, agriculture, disaster management and the establishment of communication infrastructures \cite{boursianis2022internet,kakamoukas2022fanets}.
The dynamic and distributed nature of FANETs introduces substantial security challenges, which require robust intrusion detection systems \cite{yaacoub2020security,sharma2023secure}. The increasing reliance on UAVs in critical applications underscores the unique security challenges posed by FANETs, emphasizing the need for innovative intrusion detection solutions \cite{tsao2022survey}. Unlike traditional networks, FANETs exhibit dynamic node movements, unreliable communication links, and a decentralized architecture \cite{chriki2019fanet}.

In this context, the Ad-Hoc On-Demand Distance Vector (AODV) protocol, a fundamental routing protocol in FANETs, faces challenges due to the mobility of UAVs, rendering traditional intrusion detection methods less effective \cite{amponis2021survey}. 
The AODV protocol is susceptible to various routing attacks, including blackhole, sinkhole, and flooding attacks \cite{ceviz2021analysis}. Traditional ML-based intrusion detection methods, such as Deep Neural Networks (DNN) and Convolutional Neural Networks (CNN), have been explored for their efficacy in identifying anomalous patterns \cite{nayfeh2023machine,Ouiazzane2020}. However, these methods often face challenges related to communication costs, privacy concerns, and scalability.

This paper introduces the Federated Learning-based Intrusion Detection System (FL-IDS), a novel intrusion detection approach tailored for FANETs, offering several key contributions to the field. Firstly, FL-IDS addresses privacy concerns by leveraging federated learning (FL) \cite{zhang2021survey}, a decentralized machine learning (ML) approach. This FL-based approach marks a departure from conventional intrusion detection paradigms. Secondly, our study contributes a realistic dataset based on actual FANET characteristics, providing a more authentic foundation for analysis compared to synthetic datasets used in previous research.

FL-IDS operates in a decentralized manner, enabling individual UAVs to train a global intrusion detection model collaboratively without sharing raw and sensitive data, addressing privacy concerns inherent in centralized approaches while harnessing collective intelligence. The architecture involves assigning default local models to each UAV, utilising client-specific data. After training local models, only the updated model weights are shared with a central server, which aggregates these weights to update the global model. This generated global model is responsible for detecting attacks without delaying the decisions by collecting raw data from clients, as in traditional methods. This rapid decision-making capability is crucial in promptly addressing security threats, enhancing the effectiveness of our proposed intrusion detection system. Moreover, it considerably decreases communication cost, which is very important for FANETs which have frequent link breakages due to high nodes' speeds.  

\myrev{
The role of a central server remains crucial in FANET architectures, complementing the decentralized nature of FL-IDS. While FANETs often utilize peer-to-peer connections for coordination and collaboration among UAVs, they also frequently incorporate a central server, especially for tasks like data collection and relaying information to a command control center \cite{bekmezci2013flying}. It is both practical and beneficial for certain FANET configurations due to its role in aggregating model updates rather than handling raw data transmissions. The central server is typically implemented at a Ground Base Station (GBS) or a stable UAV acting as a cluster head, as these entities possess greater computational power, stable energy supplies, and access to reliable communication infrastructure, making them well-suited for aggregating model updates. The use of a central server also provides critical benefits for resource-constrained UAVs. By offloading the task of aggregating model updates to the server, the computational and storage demands on individual UAVs are reduced. This is especially vital in FANETs, where UAVs often operate on limited battery power and computational resources.}

The novelty of FL-IDS lies in its effectiveness in detecting routing attacks, including sinkhole, blackhole, and flooding attacks, in the dynamic and decentralised FANET environment. \myrev{Additionally, the Bias Towards Specific Clients (BTSC) method, which adjusts the model's focus to prioritize clients with superior attack detection capabilities, further enhances FL-IDS's performance, even in scenarios with a low attacker ratio.} Incorporating realistic network scenarios, considering 3D node movement and local data collection, enhances the applicability of FL-IDS. A comparative analysis with traditional intrusion detection methods, including Central IDS (C-IDS) and Local IDS (L-IDS), provides valuable insights into the strengths of FL-IDS. 

\myrev{The main contributions of our study are summarized as follows:}

\myrev{
\begin{itemize}
    \item We propose a novel FL-IDS specifically designed for FANETs, addressing privacy concerns by enabling collaborative learning without sharing raw data. The decentralized operation reduces communication costs, making it suitable for the highly dynamic and resource-constrained nature of FANETs.
    \item We introduce a realistic dataset based on actual FANET characteristics, incorporating 3D node movement and local data collection. This improves the system’s applicability to real-world environments and includes critical routing attacks such as sinkhole, blackhole, and flooding attacks.
    \item The proposed system demonstrates effectiveness in detecting critical routing attacks—including sinkhole, blackhole, and flooding—while the Bias Towards Specific Clients (BTSC) method enhances FL-IDS performance even at scenarios with lower attacker ratios.
    \item A comprehensive comparative analysis with traditional IDS approaches, including Central IDS (C-IDS) and Local IDS (L-IDS), highlights the advantages of FL-IDS in effectiveness, efficiency, and privacy preservation.
\end{itemize}}

The remainder of this paper is organized as follows. In Section \ref{sec:related}, we delve into related works to contextualize our approach within existing literature. Section \ref{sec:background} provides background information on FANETs, routing attacks, and federated learning. Section \ref{sec:approach} details the proposed FL-IDS approach, explaining its architecture and functionality. The experimental results and analyses are presented in Section \ref{sec:results}. In Section \ref{discusandlimitation},  the proposed FL-IDS is discussed concerning its effectiveness, efficiency, communication cost, response time, privacy, and security. In addition, a comparative analysis is presented with C-IDS  and L-IDS across these specified criteria. Finally, Section \ref{sec:conclusion} concludes the paper, summarizing findings and outlining directions for future research.

\section{Related Work}
\label{sec:related}

This section explores recent studies and approaches relevant to intrusion detection in FANETs. We explore the latest advances, methodologies, and datasets proposed in the literature, shedding light on how researchers have tackled the challenges of FANET security. From traditional ML to emerging FL-based solutions, we summarize the key insights from prior work. This understanding helps to evaluate the landscape and guides the development of our proposed intrusion detection approach.

The use of ML (Support Vector Machines (SVM), Naive Bayes (NB), Linear Regression (LR), and Random Forest (RF)) for detecting DoS attacks against various components of a UAV is explored in \cite{baig2022securing}. The study used the DJI Phantom 4 drone dataset \cite{Dat}, which includes information collected from components such as GPS, gyro, flight controller, and battery module. While showing promising results for ML-based algorithms,\myrev{ with the best accuracy of 97.84\% achieved by RF algorithm,} the study also emphasized the need for reliable datasets to effectively train and validate ML models. Another study focusing on DoS attack detection is presented in \cite{da2022development}. This study employed the AWID2 dataset \cite{kolias2015intrusion} for training three models (XGBoost, CatBoost, and LightGBM), with tests conducted using real UAVs. The comparison of algorithms, considering factors like training time and area under the curve (AUC) metrics, revealed that LightGBM stands out among other models, \myrev{achieving an AUC of 99.83\%.} Another study \cite{shrestha2021machine} also employs traditional ML-based algorithms for detecting DDoS attacks and botnet activity in UAV networks. \myrev{Among these ML methods, the Decision Tree (DT) algorithm demonstrated exceptional performance, achieving a maximum accuracy of 99.99\%.}

Another study \cite{nayfeh2023machine} proposes a solution to detect GPS spoofing attacks categorized into two groups: static and dynamic. The dataset generated for both attacks collects GPS signal characteristics such as latitude, longitude, and time from both authentic and spoofed experiments. The study employs traditional ML algorithms such as RF, K-Nearest Neighbors (KNN), SVM, DT, and Neural Networks (NN) and uses the Spearman method for feature extraction. The results indicate that DT outperforms other algorithms, achieving a 92.36\% detection rate and a 3.7\% false positive rate. One of the main contributions of the study is providing a real-time GPS spoofing detection solution that can be seamlessly integrated with standard receivers and ubiquitous modules, requiring no hardware modifications. A multi-agent-based IDS based on DT is proposed in \cite{Ouiazzane2020}. A centrally located IDS collects data such as payload traffic, command and control traffic, and GPS data from UAVs. \myrev{The study reported achieving an accuracy of 100\%; however, }it has not been evaluated on a suitable dataset for FANETs. Hence, the authors leave the construction of a new dataset as future work.

Many of the ML-based studies mentioned earlier rely on existing datasets, which may not be well-suited for FANETs. However, in two recent studies, researchers have addressed this gap in the literature by introducing datasets specifically tailored for FANETs. The first study \cite{chulerttiyawong2023sybil} proposes an ML-based approach for detecting Sybil attacks. To facilitate this, a dataset is constructed using the OMNET++ simulator \cite{varga2010omnet}, with a particular emphasis on capturing the 3D motion and density characteristics of FANETs. This dataset includes Received Signal Strength Difference (RSS) and Time Difference of Arrival (TDoA) derived from the physical layer, utilizing ground monitoring stations. Experimental results show that this approach can detect Sybil attacks with high accuracy (91\%), with the false positive rate being less than 9\% on average.

Another attack dataset is introduced in \cite{zhai2023etd}. This study is noteworthy as it is the first to address time delay attacks in FANETs, where delays are intentionally introduced in packet transmissions to the destination. The dataset comprises latency-related data collected from networks operating under normal conditions, simulated with the ONE simulator \cite{keranen2009one}. UAVs follow pre-planned routes determined by four different routing protocols. ML techniques are employed to identify anomalies in the dataset. The K-means clustering technique is subsequently applied to distinguish between malicious and benign nodes. The study demonstrates an accuracy of over 80\% and less than 2.5\% overhead in various network settings. However, this dataset is primarily designed for pre-planned flight paths, low speed (6 m/s), and 2D movement. These specifications may not align with the requirements of some other FANET applications, as the study is explicitly tailored for search and rescue missions.

There are also studies that have explored the application of deep learning for intrusion detection in FANETs. It is shown in \cite{abu2022high} that Convolutional Neural Networks (CNNs) outperform traditional ML algorithms. The effectiveness of IDS was highlighted by an experimentally obtained accuracy of 99.50\% and a prediction time of 2.77 ms on the UAV-IDS-2020 dataset \cite{zhao2018prediction}. In \cite{Ramadan2021}, Recurrent Neural Networks (RNNs) are proposed to detect anomalies in the behavior of drones, such as unexpected changes in altitude, velocity, or trajectory. The model was trained using public datasets such as KDDCup’99 \cite{kddcup1999} and NSL-KDD \cite{5356528}. \myrev{Although it achieved an average accuracy of 95\%, these datasets are not specifically tailored to FANET environments.} The model is deployed on each UAV and GBS, allowing attacks to be detected locally and then reported to the central IDS for verification. The experimental results show that the proposed method produces better results compared to LR and KNN.

In \cite{bouhamed2021lightweight}, the Deep Reinforcement Learning (DRL) algorithm is employed to create an intrusion detection model using the CICIDS2017 dataset \cite{panigrahi2018detailed}, \myrev{achieving an accuracy of 99.70\%}. This model is deployed both on the central station and on each UAV. The approach introduces an offline learning system that automatically updates the model each time the UAV returns to its charging station, minimizing energy consumption.

Another lightweight solution based on hierarchical SVM for detecting GPS and jamming attacks is proposed in \cite{arthur2019detecting}. The IDS on the UAVs uses a DL algorithm in conjunction with the Self-Taught Learning (STL) algorithm to extract significant features from movement logs. These logs contain routing information, velocity, and GPS location data collected during flights. \myrev{While SVM achieved an accuracy of over 72\%, when combined with STL, the accuracy improves significantly, reaching over 92\%.} If an attack is detected, a Q-learning-based adaptive route learning algorithm is initiated on UAVs to return to a safe area. While the proposed approach claims to be well-suited for resource-constrained UAVs due to its utilization of DL, this claim lacks support from experimental results.

Only a few studies \cite{mowla2019federated,mowla2020afrl,da2023anomaly} have applied federated learning to detect attacks in FANETs. In \cite{mowla2019federated}, the federated learning approach is employed to detect jamming attacks using two datasets. The first dataset, consisting of 3,000 samples and 8 features such as PDR, throughput, and received signal strength indicator (RSSI), is collected using Ns-3 \cite{Ns-3}. The second dataset CRAWDAD \cite{punal2014crawdad}, originally designed for VANETs, is modified to adopt a distributed data structure that represents FANETs. Due to increased running time and communication costs when involving all clients in the training process, specific UAVs are selected for participation based on prioritization using the dumper-shaper technique. The accuracy obtained for the CRAWDAD dataset was approximately 82\%, while for the FANET dataset, it reached around 89.5\%. In contrast, traditional techniques yielded accuracies of 49.11\% and 65.62\% for the CRAWDAD and FANET datasets, respectively. In their extended study \cite{mowla2020afrl}, a reinforcement federated learning-based technique is used to identify a defense strategy in unknown areas. This technique presents a different route to the desired location by avoiding the jamming attack region using the spatial retreat technique. 

In \cite{da2023anomaly}, an IDS is proposed to detect flight anomalies and network attacks in UAV swarms. The Network Attack Detection subsystem employs a supervised approach, utilizing traditional algorithms (DT, RF, Gaussian Naive Bayes (NB), Multi-Layer Perceptron (MLP), eXtreme Gradient Boosting (XGBoost), and LightGBM) to detect network attacks (blackhole, grayhole, and flooding). For the identification of flight anomalies, an unsupervised approach is adopted, leveraging federated learning to detect GPS jamming and spoofing attacks using the UAV Attack Dataset \cite{whelan2020novelty}. Various federated learning aggregation methods (FedAvg \cite{mcmahan2017communication}, FedAdagrad \cite{reddi2020adaptive}, FedAdam \cite{reddi2020adaptive}, and FedYogi \cite{reddi2020adaptive}) are employed. Among them, FedYogi stands out for its robustness, achieving an F1-score of 0.904.

FL-based IDSs have also been proposed for IoT. An FL-based multiclass classifier is presented in \cite{CAMPOS2022108661}. The study examines three configurations as basic, balanced, and mixed, which are instances of splitting the ToN\_IoT dataset \cite{moustafa2019new} according to  IP addresses and types of attacks on IoT devices. The assessment takes into account the impact of aggregation functions such as Fed+ \cite{yu2020fed+} and FedAvg. The paper highlights the significance of an appropriate client/instance selection procedure to deal with issues in scenarios including non-independent, identically distributed, and highly skewed data. The evaluation findings emphasize the importance of this selection procedure and show that, in certain cases, using Fed + significantly improves the metrics over FedAvg, \myrev{achieving an accuracy of over 80\%}. \myrev{Another study \cite{abou2023secure} presented a secure and efficient FL approach that incorporates blockchain technology to enhance security, privacy, and trust in IDSs for IoT networks. It introduces a secure aggregation algorithm using Secure Multi-Party Computation (SMPC), a cryptographic technique that enables multiple parties to collaboratively compute a function on their private data without revealing any individual inputs. In addition, a blockchain-based reputation mechanism is used to ensure data privacy and model integrity. The proposed approach achieved an accuracy of 99\%. In another study \cite{abou2024blockchain}, blockchain and edge computing were applied in V2X networks to ensure node trustworthiness and reduce communication delays.}

In the survey \cite{ferrag2021federated}, an overview of FL-based cybersecurity studies is provided for various IoT applications, including the Internet of Vehicles, the Internet of Drones, Industrial IoT, and the Internet of Healtcare Things. The survey emphasizes the absence of FL-based datasets in the literature, which leads researchers to use and modify public datasets to mimic data federations. Even datasets collected from different environments \cite{rahman2020internet} are used for intrusion detection in IoT. However, it is imperative to utilize datasets that align with the specific characteristics of IoT or FANETs to develop more realistic intrusion detection solutions, as aimed in this study.

There are also UAV-assisted applications where UAVs are being integrated into airspace operations for different ground-based entities and networks, such as edge computing. However, the development of security strategies that involve the assistance of UAVs is still in its early stages and requires further research \cite{ALZAHRANI2020102706,9275621}.

All related studies are summarized in Table \ref{tab:rw}. As shown in the table, the majority of these studies train and evaluate their models on datasets not suitable for FANETs. Even in cases where new attack datasets for FANETs are created, their parameters are not very realistic due to simulating networks with a small number of nodes \cite{mowla2019federated}, 2D mobility models \cite{zhai2023etd}\cite{arthur2019detecting}, or nodes with low speeds suitable for MANETs and VANETs but not FANETs \cite{chulerttiyawong2023sybil}\cite{zhai2023etd}. \myrev{In contrast, FANET datasets possess distinct characteristics that set them apart from MANET and VANET datasets. These include high-speed node mobility in 3D space, frequent topology changes, and variable link reliability due to UAV dynamics. Such factors introduce significant differences in network behavior and attack patterns compared to MANETs and VANETs, which typically assume simpler 2D mobility models and more stable link conditions. Consequently, datasets designed for FANETs must incorporate realistic 3D mobility models, account for high-speed UAV nodes, and consider frequent link disruptions to ensure the reliable evaluation of IDSs in FANET environments.} Our study simulates a total of 160 networks for three attacks (sinkhole, blackhole, and flooding) with varying attacker ratios (from 5\% to 25\%) using 3D Gauss Markov Mobility (3D GM). To the best of our knowledge, this work represents the first proposal of a federated learning approach for these attacks. Although some federated learning-based studies have been proposed for jamming and spoofing attacks, their suitability has not been thoroughly evaluated in realistic network scenarios. Our study addresses this gap by exploring the use and suitability of federated learning for intrusion detection in FANETs, offering an extensive comparison with centralized and local IDS in terms of accuracy, communication cost, and privacy.

\begin{table*}[t]
  \centering
   
  \caption{Outline of Related Studies}
  \resizebox{\linewidth}{!}{ \Huge \begin{tabular}{|c|c|c|c|c|c|c|l|}
    \hline
    \text{Study} & \textbf{Year} & \textbf{Environment} & \textbf{Dataset} & \textbf{Attacks}& \textbf{Method} & \myrev{\textbf{Perf. Metrics}}&\textbf{Pros (+) \& Cons (-)}\\
    \hline    \cite{Ouiazzane2020}&2020&FANETs&CICIDS2017 \cite{panigrahi2018detailed}&Brute force, DoS, BotNet, Port Scanning&ML&Accuracy&+Multi-agent-based IDS is proposed\\
    & & & & SQL Injection, XSS, Heartbleed&&&-Dataset is not suitable for FANETs\\
      &  &  &  && && -Semi-supervised machine learning methods are not evaluated\\
    \hline
    \cite{shrestha2021machine}&2021&Cellular Connected UAV Networks&CSE-CIC-IDS2018 \cite{sharafaldin2018toward}&Brute force, DoS, DDoS&ML&Accuracy, Precision& +A study on cellular-based study\\
    
    & & & & BotNet, Web Attack, Infiltration&&Recall, F1-score, FNR&  -The dataset is not suitable for FANETs and 5G networks \\
   
    \hline

  \cite{baig2022securing}&2022&UAVs&DJI Phantom 4 \cite{Dat}&DoS attacks&ML&Accuracy, Precision& +Evaluates various ML algorithms\\
  & & & & &&Recall, AUC& -Focused on sensor-based attacks rather than FANETs-specific attacks\\
  \hline
   \cite{da2022development}&2022&UAVs& AWID2 \cite{kolias2015intrusion} (for training and testing in simulation) &DoS attacks&ML&Accuracy, Precision& +Various ML-based algorithms are assessed with real UAVs \\
    & & & individual UAV data transmission (for real testing) & &&Recall, F1-score, FPR& +Providing hyperparameter optimization using Bayesian algorithm\\
  & & &(speed, pitch and temperature, etc.) & &&& -Dataset is not suitable for FANETs \\
  \hline
   \cite{abu2022high}&2022&FANETs&UAV-IDS-2020 \cite{zhao2018prediction}& attacks for unauthorized access in WLAN &ML&Accuracy, Precision, TPR& +Data collected from real UAVs were used. \\
  & & & & &&Recall, F1-score, FPR& -Dataset is not suitable for FANETs \\
  \hline
% & & & & && - ...the performance on detection each attack is not discussed\\
   \cite{nayfeh2023machine}&2023&UAVs& UAV dataset & GPS Spoofing &ML&Accuracy, Precision, TPR& +Various ML-based algorithms are tuned and evaluated\\
    & && (a real UAV was used)  & &&Recall, F1-score, FPR&\\
  \hline
   \cite{chulerttiyawong2023sybil}&2023&FANETs&a FANET dataset &Sybil&ML&Accuracy& +A new dataset is created \\
    & & & (100 nodes, 3D INET’s MassMobility) & &&TPR& -It is evaluated on a small area 
    with nodes having low speeds (10-20 m/s) \\
    & & & & && -High false positive rate\\
  \hline
  \cite{zhai2023etd}&2023&FANETs&a FANET dataset &Time Delay&ML&Accuracy& +A new dataset is created by using 4 different routing protocols\\
    & & &  (13, 24 nodes and 2D MapRouteMovement) & &&& - 2D mobility is considered\\
    & &  & & &&& -Use of pre-planned route applications \\
  \hline
    \cite{arthur2019detecting}&2019&FANETs&a FANET dataset collected from nodes  &Spoofing, Jamming&DL&Accuracy& +It is claimed to be lightweight\\
    &&&(20 nodes, Random Way Point Mobility Model ) &  &&& -The resource consumption is not given in the results \\
    &&&&  && &-The dataset is collected using 2D mobility model\\
    \hline
\cite{Ramadan2021}&2021&FANETs&KDDCup'99 \cite{kddcup1999}&DoS, DDoS, Probe, R2L, U2R&DL&Accuracy, Precision& +Both local and central IDSs participate in detection\\
     & & &NSL-KDD \cite{5356528}&Fuzzer, Analysis, Backdoors, Exploits &&TPR, F1-score& -The datasets are not suitable for FANETs\\
    & & &UNSW-NB15  \cite{moustafa2015unsw} &Worms, Shell Code, Reconnaissance &&&  \\
    & & &WSN-DS \cite{almomani2016wsn}&Password, XSS, Injection, Scanning&&&\\
    & & &CICIDS2017 \cite{panigrahi2018detailed}&&&&\\
    & & &TON\_IoT \cite{moustafa2019new}&&&&\\
    \hline

    \cite{bouhamed2021lightweight}&2021&FANETs&CICIDS2017 \cite{panigrahi2018detailed} &Brute force, DoS, DDoS&DL&Accuracy, Precision& +Offline-learning is proposed, where UAVs can update models during charging\\
    &&& &  BotNet, Web Attack, Infiltration&&Recall, F1-score& -The dataset is not suitable for FANETs \\
    \hline
       \cite{mowla2019federated}& 2019 &  FANETs & CRAWDAD \cite{punal2014crawdad} for VANETs & Jamming & FL&Accuracy& +A client group prioritization technique is proposed\\
    &  &  & a FANET dataset  (6 nodes, 3D GM  Mobility Model)  &  &&Running Time& -The effect of mobility on client group prioritization is not discussed \\
      &  &  &  && && -It is evaluated on very small networks (6 \& 3 clients) \\
    \hline
    \cite{rahman2020internet} & 2020 & IoT & NSL-KDD \cite{5356528} & DoS, U2R, R2L, Probing attacks for training & FL&Accuracy& +Data distribution is explored for different types of attacks \\ 
    & & & & 17 different attack types for testing &&& -The dataset is not suitable for IoT\\
    
    \hline
    \cite{ferrag2021federated} & 2021 & IoT & Bot-IoT \cite{koroniotis2019towards} & DoS, DDoS, Theft, Reconnaissance & FL&Accuracy, Precision& +Different learning algorithms are evaluated \\
    &  &  & MQTTset \cite{vaccari2020mqttset} & Specific Attacks against MQTT & &Recall, F1-score&+The dataset is collected from heterogeneous devices\\
    &  &  & TON\_IoT \cite{moustafa2019new} & Password, Backdoor, XSS, Injection, Scanning & &&-Data distribution is mimicked for FL\\
    \hline
   
      \cite{CAMPOS2022108661}&2022&IoT& ToN\_IoT\cite{moustafa2019new} &Backdoor,Dos,DDos,Injection,MITM&FL&Accuracy, Precision& +Realistic dataset with consideration of non-independent and identically distributed data   \\
     & & & &Password,Ransomware,Scanning,XSS &&Recall, F1-score, FPR& -Resource consumption is not examined\\
  
    \hline
        \myrev{\cite{abou2023secure}}&2023&IoT& UNSW-NB15 \cite{moustafa2015unsw}  & Fuzzers, Analysis, Backdoors & FL &Accuracy, Precision& +Combines FL with Blockchain to ensure privacy, security, and trust\\
     && & & DoS, Exploits, Generic &Blockchain&Recall, F1-score&+Uses SMPC for secure aggregation, preserving data privacy   \\
    & & &  & Reconnaissance, Shellcode and Worms &  &Training Time& -Increased computational and communication overhead\\
       & & &  &  & & & -Blockchain implementation can be complex and resource-intensive\\
  
    \hline
     \cite{da2023anomaly}&2023&UAV swarm & WSN-DS \cite{almomani2016wsn}  & Blackhole, Grayhole, Flooding & ML &F1-score, AUC& +Various ML-based algorithms are evaluated\\
     && & UAV attack dataset \cite{whelan2020novelty} & Ping DoS &&Training Time&-A significant percentage of the normal class is classified as grayhole  \\
    & & &(6 real UAVs)  & GPS Spoofing, Jamming & FL && -Blackhole attacks ands normal traffic falsely detected as grayhole \\
  \hline 
  \textbf{Our} & 2025 & FANETs & a FANET 
  dataset  & Sinkhole, Blackhole, Flooding & FL &Accuracy, Recall& +A new dataset is created\\
     \textbf{Study} & &  & (50 nodes, 3D GM Mobility Model,  &  &  & FPR&+Extensive simulations are carried out \\ 
     & & & attacker ratios: 5\%, 10\%, 15\%, 20\%, 25\%, &  &  && +Rigorous comparison with centralized and local intrusion detection\\
     & & &average speed: 100 m/s)   &  &  & &-Communication cost during training\\\hline
  \end{tabular}}
  
   \label{tab:rw}
\end{table*}

\section{Background}
\label{sec:background}

This section provides essential context for our study on intrusion detection in FANETs. We introduce the concept of federated learning as a promising approach to enhance security. Additionally, we explore the key components of FANETs, such as the AODV routing protocol and common attacks like sinkhole, blackhole, and flooding attacks. 

\subsection{Federated Learning}
The utilization and processing of the data collected from distributed devices pose challenges for traditional centralized IDSs. Firstly, there is the violation of privacy concern related to data sharing. In the conventional approach, the central server has access to all private client data to train a global model, which is then shared with all clients \cite{agrawal2022federated}. However, this raises privacy issues, as the data may contain private information that clients prefer not to share with third parties. Secondly, in the context of FANETs characterized by high node speeds and frequent link breakages, relying on communication among IDS clients may not be preferable due to potential disruptions and unreliability in data transmission. This exacerbates the challenges posed by dealing with large volumes of data in a distributed environment, resulting in increased computational and communication costs, ultimately causing latency in a central IDS.
Lastly, energy consumption becomes a significant concern for resource-constrained devices, and both clients and central systems might require high energy consumption in order to obtain, store, process the local and the aggregated data respectively. Solutions that consider energy consumption must be designed for FANETs, particularly due to the dependence of mini UAVs on low-capacity batteries for power \cite{Ullah2017}.

\begin{figure}[t]
 \centering
\includegraphics[width=\textwidth]{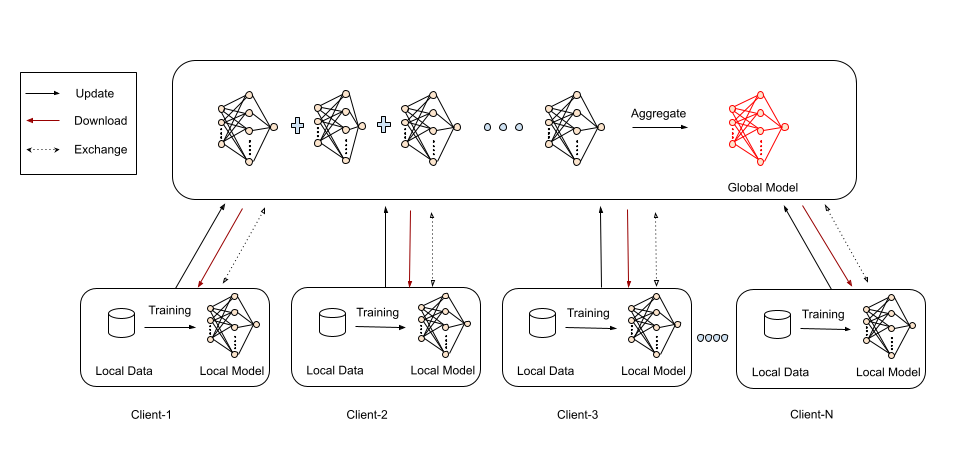}
\caption{Architectural Overview of of Federated Learning} \label{Fig1}
\end{figure}

Federated Learning (FL) was proposed in \cite{mcmahan2017communication} mainly to overcome privacy issues as an alternative to traditional methods. In FL, data is distributed across multiple devices or servers, and the model is trained locally on each of these devices. Figure \ref{Fig1} shows one of the well-known methods in FL called Federated Averaging (FedAvg). In this method, after training the models locally, the updates are sent back to a central server, where they are aggregated, and the global model is updated. For example, parameters such as weight averages of models trained on clients can be sent to the central server for aggregation, and a single federated model is returned to clients. This process can be repeated during a certain round until the desired accuracy value is achieved or for a certain duration. Another method, Federated Learning with Differential Privacy, involves adding noise to the model updates to protect the privacy of local data. This noise, added using differential privacy techniques, ensures the protection of local data privacy while still allowing effective training of the global model. Another privacy and security-focused model is Secure Federated Learning. This method uses cryptographic techniques like homomorphic encryption and secure multi-party computation, allowing devices to collaboratively train a model without revealing their local data or model parameters to other devices or the central server. These are just a few of the many methods used in federated learning. The choice of FL method depends on the specific use case, the type of data, and the privacy and security requirements of the participants and applications. In this study, we employ aggregated federated averaging due to commonly used aggregation strategy that offers fast convergence, simplicity, and wide applicability. 

The general steps for implementing FL-based IDSs in this study are listed below.
\begin{itemize}
    \item Initialization: The central server distributes an initial model to the participating devices or servers.
    \item Training: Each device or server independently trains the model using its local data. This local training may involve multiple rounds, during which the device or server computes an update to the model based on its local data.
    \item Aggregation: Updates from all participating devices or servers are sent back to the central server and combined to produce a new version of the model.
    \item Iteration: The training, updating and aggregating process is repeated until the model reaches an acceptable level of accuracy.
\end{itemize}

FL is a promising approach for our study, driven by several key advantages. In particular, FL significantly enhances efficiency by mitigating the need to transmit large amounts of data to a central location for processing, resulting in savings in both time and bandwidth. This reduction in the cost of data storage and processing renders FL a cost-effective solution, particularly beneficial for resource-constrained devices and environments.

 Moreover, its inherent scalability and flexibility make FL adaptable to a wide range of AI tasks. Notably, FL prioritizes privacy, a paramount concern in today's data-driven landscape. Additionally, FL's ability to handle missing data and adapt to data drifts (common occurrences in real-world applications) ensures the development of robust models \cite{che2023multimodal}. Consequently, our decision to utilize FL aligns seamlessly with our overarching aim: to harness the power of ML while upholding data privacy and security. This proposition is particularly appealing for IDSs seeking effective, privacy-preserving, and distributed AI solutions.

\subsection{Ad hoc On-Demand Distance Vector (AODV)}

The AODV routing protocol is a reactive routing protocol originally designed for MANETs. Unlike proactive routing protocols that maintain a complete network topology, AODV creates routes only when a node needs to transmit data to another node. When a node intends to transmit data to another node beyond its immediate range, it initiates a route discovery process by broadcasting a route request (RREQ) packet. This RREQ packet propagates through the network until it reaches either the destination node or a node that has a sufficiently fresh route to the destination. Subsequently, the route is established by sending a route reply (RREP) packet back to the source node. During this process, when the destination node or an intermediate node processing an active route to the destination in its routing table receives an RREQ packet, it sends a unicast RREP packet to the source node. The source node then selects the shortest and most up-to-date route based on the minimum hop count and the highest destination sequence number, respectively. This route selection ensures the transmission of data through the shortest and most up-to-date route available. 

AODV also incorporates a route maintenance mechanism to effectively handle link failures and route breakdowns. This process begins with the transmission of a Route Error (RERR) message. Upon detecting a link break or node failure, the node generates and sends the RERR message to the nodes that possess a route to the unreachable destination. This action promptly invalidates the affected routes. Subsequently, the affected node initiates a new route discovery process to find an alternative path to the destination. AODV is highly regarded in the context of FANETs due to its low overhead and its rapid adaptability to network changes. 

\subsection{Attacks}

In this section, we introduce attacks on AODV, which will serve as the basis for training and testing the proposed approach in this study. 

\subsubsection{Sinkhole Attack}
In this attack, the attacker falsely claims to offer a shorter route to the destination node when a source node initiates a route discovery mechanism \cite{Daniel2014}. If the source node selects this deceptive route, the attacker node can eavesdrop on all network traffic between the source and destination nodes, effectively executing a sinkhole attack \cite{ceviz2021analysis}. This initial attack is typically performed as a preliminary step to subsequent actions, including data packet dropping and modification.

\subsubsection{Blackhole Attack}

A blackhole attack combines elements of sinkhole and dropping attacks, enabling the attacker to selectively discard packets destined for a specific destination. The attacker can choose to drop all received packets to disrupt network communication or employ random dropping to evade detection. Similarly to a sinkhole attack, the blackhole attacker initially advertises itself as having the best route to the desired destination, attracting network traffic towards it. Subsequently, additional attacks, such as data modification and packet dropping may be executed. As a result, a blackhole attack can severely disrupt network communication and lead to increased energy consumption.

\subsubsection{Flooding Attack} A flooding attack, a type of Denial-of-Service (DoS) or Distributed Denial-of-Service (DDoS) attack, overwhelms a destination or network with a high volume of traffic. This results in the target slowing down or becoming unresponsive, disrupting normal operation and rendering it inaccessible to intended users.

Flooding attacks can manifest in various forms, including the transmission of excessive routing control and data packets. Attackers often exploit vulnerabilities in routing protocols, targeting route discovery and neighbor discovery mechanisms. For instance, an attacker may flood the network with numerous Hello packets, exploiting protocols that identify nearby neighbors. A type of flooding attack, known as an \textit{ad hoc flooding attack}, exploits the route discovery mechanism of AODV by broadcasting numerous RREQ packets at regular intervals. These packets may potentially request routes for non-existent nodes, leading to the consumption of network and node resources. This, in turn, results in congestion and isolation of nodes, effectively achieving the attacker's goal.

\section{The Proposed Approach}
\label{sec:approach}

In this section, we detail the implementation of our attacks, the creation of datasets, and the application of both traditional and federated learning methods to evaluate the proposed IDSs.

\subsection{Network and Attack Settings}

We conducted the three distinct attacks, introduced earlier, namely sinkhole, blackhole and ad hoc flooding, against the AODV routing protocol within FANETs, utilizing the Ns-3 simulator \cite{Ns-3}. Each network consisted of 50 mobile nodes to simulate the multi-hop characteristics of ad hoc networks. Furthermore, an immobile Ground Base Station (GBS) node was positioned at the simulation center due to its utility in various applications, including disaster response, agricultural monitoring, and military operations \cite{zhi2020security,bai2023towards}. The GBS serves as a central hub for receiving, processing, and analyzing data collected by the UAVs. With this setting, our scenario simulates both the Air-to-Air (A2A) communication channel and the UAV to GBS link as Air-to-Ground (A2G) channel communication\cite{ALZAHRANI2020102706,cui2021channel}.

The 3D natural flight of UAVs was emulated with the 3D Gauss Markov (GM) mobility model to provide a realistic approach \cite{6127781}. This model mimics real-world scenarios, producing smoother turns and predictable movement paths instead of entirely random motion. Within the network, 10 source nodes and 10 destination nodes communicate randomly, with any remaining nodes potentially serving as relay nodes. Data collection and transmission to the GBS are the responsibilities of the destination nodes, beginning at the 10th second and continuing every second until the end of the simulation.

The attack nodes were randomly selected from the non-source or non-destination nodes. For each attacker ratio, ranging from 5\% to 25\%, the attacker nodes were randomly chosen. These same attacker nodes were used consistently in each type of attack. The network simulation parameters are listed in Table \ref{table:2}. These parameters were carefully chosen to align with real-life scenarios, considering factors such as 3D natural flight patterns, area selection, and an adaptable routing protocol for dynamic FANET movements. More details on parameter selection can be found in our previous study \cite{ceviz2023survey}.

\begin{table}[h]

\caption{Network Simulation Parameters}
\label{table:2}
\centering

\resizebox{0.5\columnwidth}{!}{%
\begin{tabular}{|c |c |} 
 \hline
 Parameters & Values \\ [0.5ex] 
 \hline\hline
 Routing protocol & AODV \\ 
 \hline
 MAC protocol & IEEE 802.11b\\
 \hline
 Channel characteristics & YansWifiChannel, PropagationLossModel\\
 \hline
 Simulation time & 1800 seconds\\
 \hline
 Simulation area &  12000 m x 12000 m x 300 m\\
 \hline
Number of nodes & 50\\
\hline
Average speed & 100 m/s \\
\hline
Transmission range & 250 m\\
\hline
Traffic type & UDP with 10 connections \\
\hline
Packet size & 512 bytes\\
\hline
Packet rate & 1/s \\
\hline
Bandwidth & 11 Mbps \\
\hline
Attacker ratio & no attack, 5\%, 10\%, 15\%, 20\%, 25\%\\
\hline
Mobility model & 3D GM \\
\hline
Bounds for GM Mobility & X: [0; 12000],
Y: [0; 12000], Z: [0; 300]\\
\hline
$\alpha$ for GM Mobility & [0.25-0.7]\\ [1ex]
 \hline
\end{tabular} }
\end{table}

\begin{table}[t]
\caption{Dataset Information}

\label{table:3}
\centering
\resizebox{0.5\columnwidth}{!}{%
\begin{tabular}{|c |c |} 
 \hline
Attributes & Values \\ [0.5ex] 
 \hline \hline
 Number of network simulations & 160 \\ 
 \hline
 \hline
 Number of samples & 5,492,700 \\ 
 \hline
 Number of features & 31\\
 \hline
 Labels & Normal, Attack\\
 \hline
 Attack Type & Sinkhole, Blackhole, Flooding\\
 \hline
Number of nodes & 51\\
\hline
Data collection period & every 5s\\
 \hline
\end{tabular}}
\end{table}

\begin{figure}[h]
\centering
\includegraphics[width=.8\linewidth]{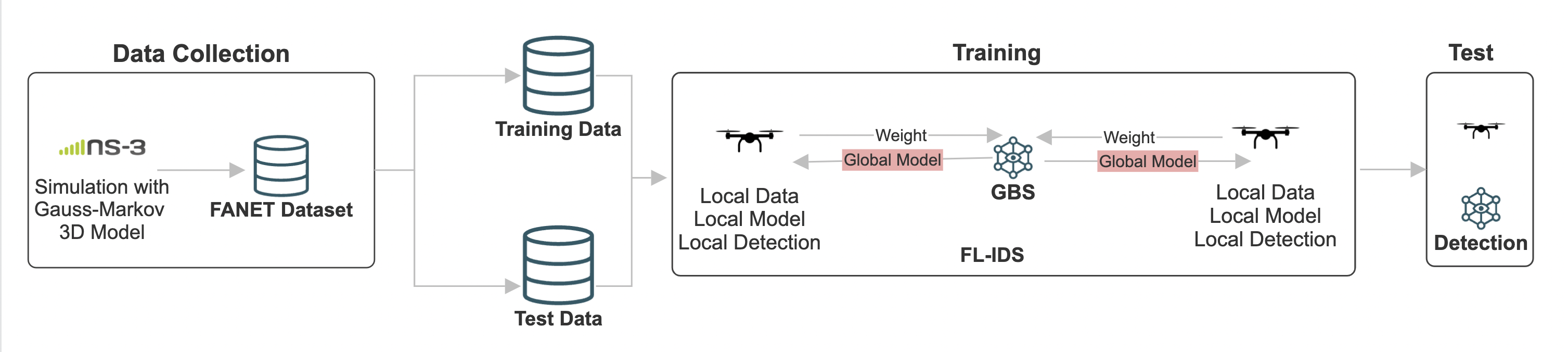}
\caption{Conceptual Schema of the Proposed Approach} \label{fig:network_arc}
\end{figure}
Here, the specifics of each attack and their implementation in our study is given in details. 

\textit{Sinkhole Attack}: In this attack, upon receiving an RREQ packet from the source node, the attacker generates a fake RREP packet with a higher destination sequence number and claims to be just one hop away from the destination. This guarantees the selection of the attacker's route, effectively attracting network traffic to itself.

\textit{Blackhole Attack}: The blackhole attack combines elements of the sinkhole and dropping attacks. Initially, the attacker establishes a route, attracting traffic as described in the sinkhole attack. Subsequently, it indiscriminately drops every packet it receives, thus disrupting communication between endpoints. However, the attacker's position is critical in dropping attacks. If it is located on a passive route, it has no impact on network performance. Due to this and the need for knowledge from other layers to differentiate it from dropping caused by wireless links, dropping attacks alone were not used in the simulations.

\textit{Flooding Attack}: In this scenario, an attacker node periodically sends RREQ packets to randomly selected destinations within the network. The chosen destination node receives multiple sequential (= 10) RREQ broadcast messages, with this process repeated every 3 seconds throughout the simulation.

To ensure a comprehensive evaluation, we created ten different network topologies with diverse network and mobility patterns, initially without any attack implementations. Subsequently, we introduced malicious nodes into these networks, enabling us to simulate a wide range of attack scenarios. We conducted 10 simulations for each attack, each with a different attacker ratio, ranging from 5\% to 25\%. Additionally, we ran 10 simulations without any attacks, totaling 160 simulations. Data were collected from each simulation over a period of 1800 seconds.

\subsection{Federated IDS}

The study introduces a Federated Learning-based Intrusion Detection System (FL-IDS) to detect network attacks in FANETs and compares it to traditional methods. \myrev{Figure \ref{fig:network_arc} illustrates the main steps of the proposed approach.} The first step is data collection, during which we collected \myrev{all} the features introduced in \cite{sen2011evolutionary} from each node. The extracted features cover a wide range of characteristics, including those related to mobility, as well as those associated with AODV control messages and data packets, as described in \cite{sen2011evolutionary}. For example, certain features provide information on the frequency of forwarding, sending, and receiving routing protocol control packets (RREQ, RREP, and RERR). \myrev{Additionally, some features reflect changes in the routing table, which can be attributed to mobility. Furthermore, features such as variations in the number of neighbors offer direct indications of mobility patterns.} These features were collected from each node at 5-second intervals.
The dataset information is summarized in Table \ref{table:3}. Please note that while the frequency of data collection positively affects the performance of the system, it also increases resource consumption. Further exploration of trade-offs in the selection of this parameter can be conducted in future studies.

To facilitate model training and testing, the dataset was split into 80\% for training and 20\% for testing, with a standard scaler applied. All IDS models employ Deep Neural Networks (DNN) and Convolutional Neural Networks (CNN). Three types of IDS are considered: Central IDS (C-IDS), Local IDS (L-IDS), and FL-IDS. The performance of traditional and federated learning methods on IDS is assessed in terms of accuracy and is compared across criteria such as effectiveness, efficiency, communication, privacy, and security in subsequent sections.

The following subsections detail the specifics of each IDS.

\begin{itemize}
    \item \emph{C-IDS:} C-IDS collects data (local features) from all nodes and aggregates them to train a single, centralised model. The model is centrally located at GBS.  
    \item \emph{L-IDS:} In this approach, each node is individually trained with its own local data to create its own model. Then each local model is deployed to all UAVs. Please note that the evaluation result presented is the average of all L-IDS results.
    \item \emph{FL-IDS:} In the proposed method, each UAV is initially assigned a default local model to incorporate client-specific data in the IDS. Each client/node trains the default model with their local data, creating new weights. After this process, only the fresh model's weights are shared with the central server located at the GBS. The server aggregates the weights of each client, updating the global model with the average of these weights using the FedAvg method \cite{mcmahan2017communication}. The updated global model's weights are then sent back to each client for further training iterations, and these processes continue until the defined epoch. 

    \myrev{To address the dynamic and heterogeneous nature of FANETs, we employed a methodology called \textbf{\emph{Bias Towards Specific Clients (BTSC)}} to enhance the performance of our FL-IDS. BTSC prioritizes clients who demonstrate superior attack detection capabilities by assigning higher weights to their model updates during the federated aggregation process. Clients are ranked on the basis of their detection performance metrics, ensuring that the most effective models contribute more significantly to the global model. The top 20\% of clients, identified from previous experiments based on their high detection accuracy, are selected for aggregation.}
    
    \myrev{ To mitigate the risk of overfitting to these high-performing clients, the global aggregated model is evaluated on a comprehensive data set that includes data from all clients, both contributing and non-contributing. This evaluation process ensures that the model is effectively generalized across diverse data distributions, maintaining robust detection performance without bias toward specific clients. Our results confirm that the approach improves the effectiveness of the global model while preserving its generalizability.}
 \end{itemize}

These approaches have a number of parameters that need to be tuned for optimal performance. To find the best parameters for the models, the Grid Search method is employed. This method exhaustively searches for a predefined set of hyperparameters and selects the combination that produces the best performance. The best parameters selected for optimal performance are shown in Table \ref{table:4}.

The performance of models is evaluated on the test set using the following metrics: accuracy, detection rate (DR) and false positive rate (FPR). The final results are obtained by averaging the results of the method applied to different topologies for each IDS.

\myrev{This novel, distributed approach is specifically designed to align with the unique characteristics of FANETs, such as their high mobility, decentralized architecture, and dynamic network topologies. Using federated learning, FL-IDS reduces communication costs and preserves privacy by sharing only model weights instead of raw data.} This allows the central server to aggregate model weights without directly accessing raw data from individual nodes, addressing the disadvantages of both C-IDS and L-IDS, such as increased communication costs and privacy violations. By sharing only the model weights, we effectively mitigate privacy concerns while enabling collaborative analysis and intrusion detection in a secure manner. This approach is particularly well-suited for distributed systems like FANETs. \myrev{Our method is applied to routing attacks, using a dataset specifically tailored to FANETs that includes features like 3D mobility patterns and realistic data collection, ensuring a practical and accurate evaluation of the system.} It is worth noting that most studies in this field \cite{ferrag2021federated} do not reflect the distribution of client-specific data in the real world due to the lack of FL-specific datasets.

\begin{table}[h!]
\caption{Deep Learning Parameters}
\small
\label{table:4}
\centering
\resizebox{0.5\columnwidth}{!}{%
\begin{tabular}{|c |c |c |} 
 \hline
 Algorithms &Parameter& Value \\ [0.5ex] 
 \hline\hline
 DNN & Number of neurons &8 - 16 \\ 
 &Number of hidden layers&2\\
 \hline
 CNN & Number of nodes&8 - 16\\
 &Number of hidden layers&2\\
 &Convolutional layers& 1 Conv1D\\
 &Pooling layers&1 MaxPool1D\\
 &Dropout& 0.1 \\
 &Kernel size&3\\
 &Filters&22\\
 \hline
 Used in both & Optimizer&SGD\\
 &Learning rate &0.01\\
 &Loss function &binary\_crossentropy\\
 &Batch size &32\\
 &Activation function &ReLu\\
 &Classification function &sigmoid\\
 &Number of local epoch&1\\
 &Number of global epoch&100\\
 \hline

\end{tabular}}
\end{table}

\section{Experimental Results}
\label{sec:results}

In this section, we present and analyze the results of our experiments, providing a comprehensive evaluation of the proposed FL-IDS. The performance of FL-IDS is compared with traditional intrusion detection methods, namely Central IDS (C-IDS) and Local IDS (L-IDS), across various attack scenarios, including sinkhole, blackhole, and flooding attacks, under different attacker ratios in the FANET. In addition, we introduce and discuss the effectiveness of the Bias Towards Specific Clients (BTSC) method, which further refines FL-IDS performance. The experimental results offer valuable insights into the strengths and capabilities of FL-IDS in addressing the security challenges posed by dynamic and decentralized UAV networks.

\subsection{Classifier: DNN vs CNN}  

In this subsection, the performance of the DNN and CNN algorithms is compared for all types of simulated attacks. The results are shown in Tables \ref{table:6}, \ref{table:5}, \ref{table:7} for sinkhole, blackhole, and flooding attacks, respectively. Across all IDSs, CNN outperforms DNN. The same trend is observed for the proposed FL-based approach. In FL-IDS, while CNN is better than DNN across all attacker ratios, the gap between the two algorithms widens as the attacker ratio increases. Since CNN consistently outperforms DNN, we will present only the CNN results in the subsequent sections. 

\begin{table}
\centering
\caption{Accuracy of DNN and CNN in Detecting Sinkhole Attack}
\label{table:6}
\resizebox{0.5\linewidth}{!}{%
\begin{tabular}{|c|l|l|l|l|l|l|} 
\hline
                    & \multicolumn{2}{c|}{C-IDS} & \multicolumn{2}{c|}{L-IDS} & \multicolumn{2}{c|}{FL-IDS}  \\ 
\hline
Attacker Ratio (\%) & DNN     & CNN                    & DNN     & CNN                  & DNN     & CNN               \\ 
\hline
5                   & 69.03\% & 82.85\%                & 62.39\% & 72.99\%              & 64.00\% & 70.44\%           \\ 
\hline
10                  & 80.21\% & 93.33\%                & 67.54\% & 81.19\%              & 82.00\% & 89.19\%           \\ 
\hline
15                  & 81.53\% & 95.21\%                & 70.54\% & 85.40\%              & 89.41\% & 94.07\%           \\ 
\hline
20                  & 92.78\% & 98.06\%                & 73.06\% & 88.28\%              & 89.33\% & 97.41\%           \\ 
\hline
25                  & 99.31\% & 99.03\%                & 75.33\% & 90.27\%              & 91.63\% & 97.70\%           \\
\hline
\end{tabular}%
}
\end{table}

\begin{table}
\centering
\caption{Accuracy of DNN and CNN in Detecting Blackhole Attack}
\label{table:5}
\resizebox{0.5\linewidth}{!}{%
\begin{tabular}{|c|l|l|l|l|l|l|} 
\hline
               & \multicolumn{2}{c|}{C-IDS} & \multicolumn{2}{c|}{L-IDS} & \multicolumn{2}{c|}{FL-IDS}  \\ 
\hline
Attacker Ratio (\%) & DNN     & CNN                    & DNN     & CNN                  & DNN     & CNN               \\ 
\hline
5                   & 73.89\% & 83.68\%                & 62.27\% & 73.85\%              & 64.67\% & 65.04\%           \\ 
\hline
10                  & 73.61\% & 93.26\%                & 66.73\% & 81.50\%              & 81.56\% & 90.00\%           \\ 
\hline
15                  & 76.73\% & 96.21\%                & 70.30\% & 85.87\%              & 89.04\% & 95.10\%           \\ 
\hline
20                  & 91.32\% & 98.33\%                & 72.77\% & 88.54\%              & 96.67\% & 99.04\%           \\ 
\hline
25                  & 92.18\% & 98.89\%                & 75.08\% & 90.35\%              & 97.48\% & 99.26\%           \\
\hline
\end{tabular}%
}
\end{table}

\begin{table}[!h]
\centering
\caption{Accuracy of DNN and CNN on Detecting Flooding Attack}
\label{table:7}
\resizebox{0.5\linewidth}{!}{%
\begin{tabular}{|c|l|c|l|c|l|c|} 
\hline
                & \multicolumn{2}{c|}{C-IDS} & \multicolumn{2}{c|}{L-IDS} & \multicolumn{2}{c|}{FL-IDS}  \\ 
\hline
Attacker Ratio (\%) & DNN     & CNN                    & DNN     & CNN                  & DNN     & CNN               \\ 
\hline
5                   & 76.80\% & 80.46\%                & 65.89\% & 75.27\%              & 74.78\% & 76.12\%           \\ 
\hline
10                  & 98.47\% & 99.93\%                & 79.04\% & 87.41\%              & 76.52\% & 97.78\%           \\ 
\hline
15                  & 94.44\% & 99.65\%                & 83.27\% & 99.35\%              & 84.87\% & 98.30\%           \\ 
\hline
20                  & 99.17\% & 99.51\%                & 84.75\% & 99.05\%              & 91.19\% & 99.26\%           \\ 
\hline
25                  & 98.68\% & 99.58\%                & 85.57\% & 99.07\%              & 93.70\% & 99.33\%           \\
\hline
\end{tabular}%
}
\end{table}

\subsection{Effectiveness: Central, Local vs Federated IDS} 

In this subsection, we compare three intrusion detection approaches: Central IDS (C-IDS), local IDS (L-IDS), and federated IDS (FL-IDS) for each type of attack. We first discuss the results separately for each attack type and then provide a general discussion.

\subsubsection{Sinkhole Attack} 

\begin{figure*}[h!]
  \centering
  % First row: Sinkhole and Blackhole side by side
  \begin{subfigure}[b]{0.49\linewidth}
    \centering
    \includegraphics[width=.9\linewidth]{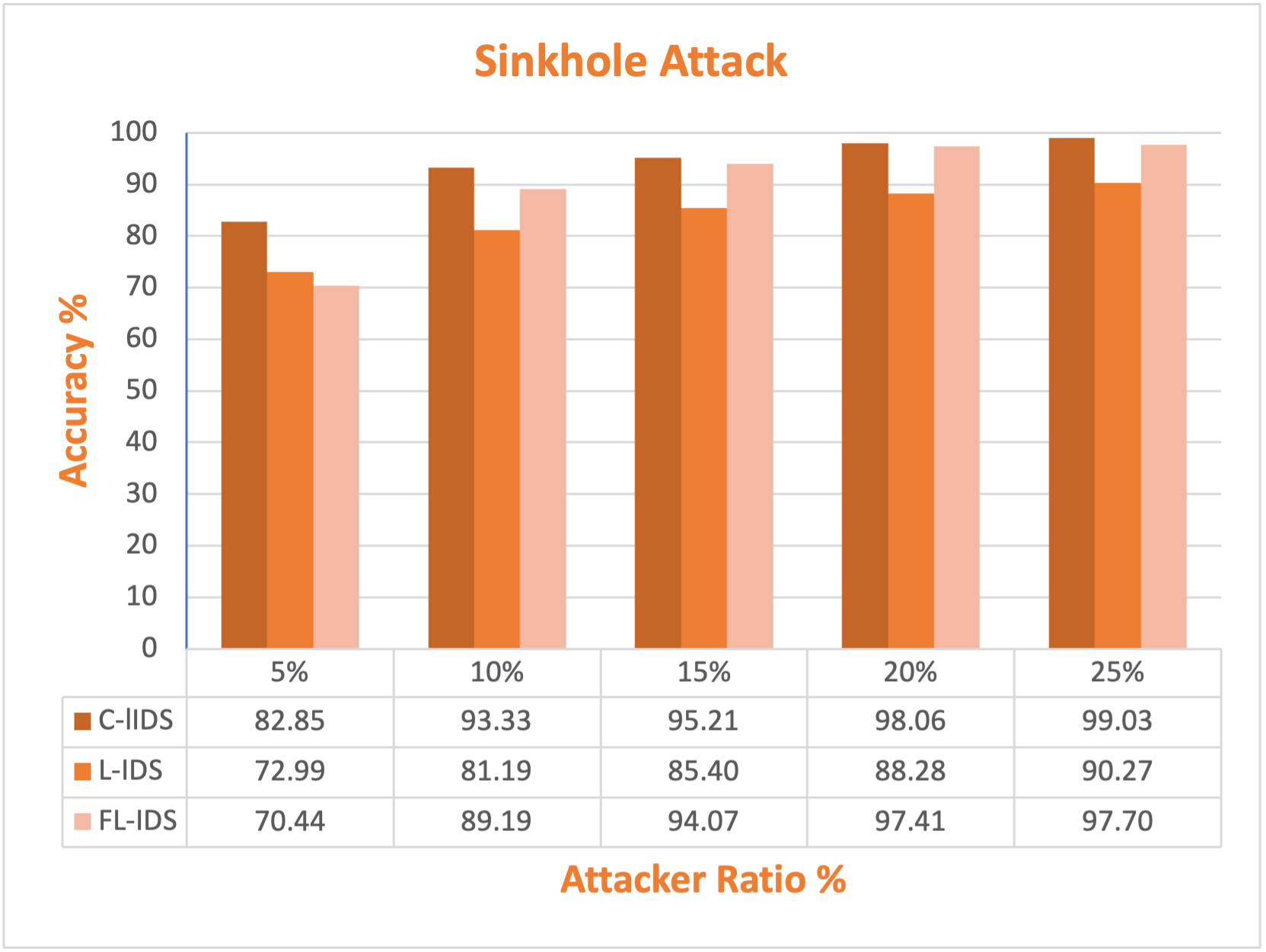}
    \caption{Sinkhole Attack}
    \label{fig:sinkhole} % Properly associated with the subcaption
  \end{subfigure}\hfill
  \begin{subfigure}[b]{0.49\linewidth}
    \centering
    \includegraphics[width=.9\linewidth]{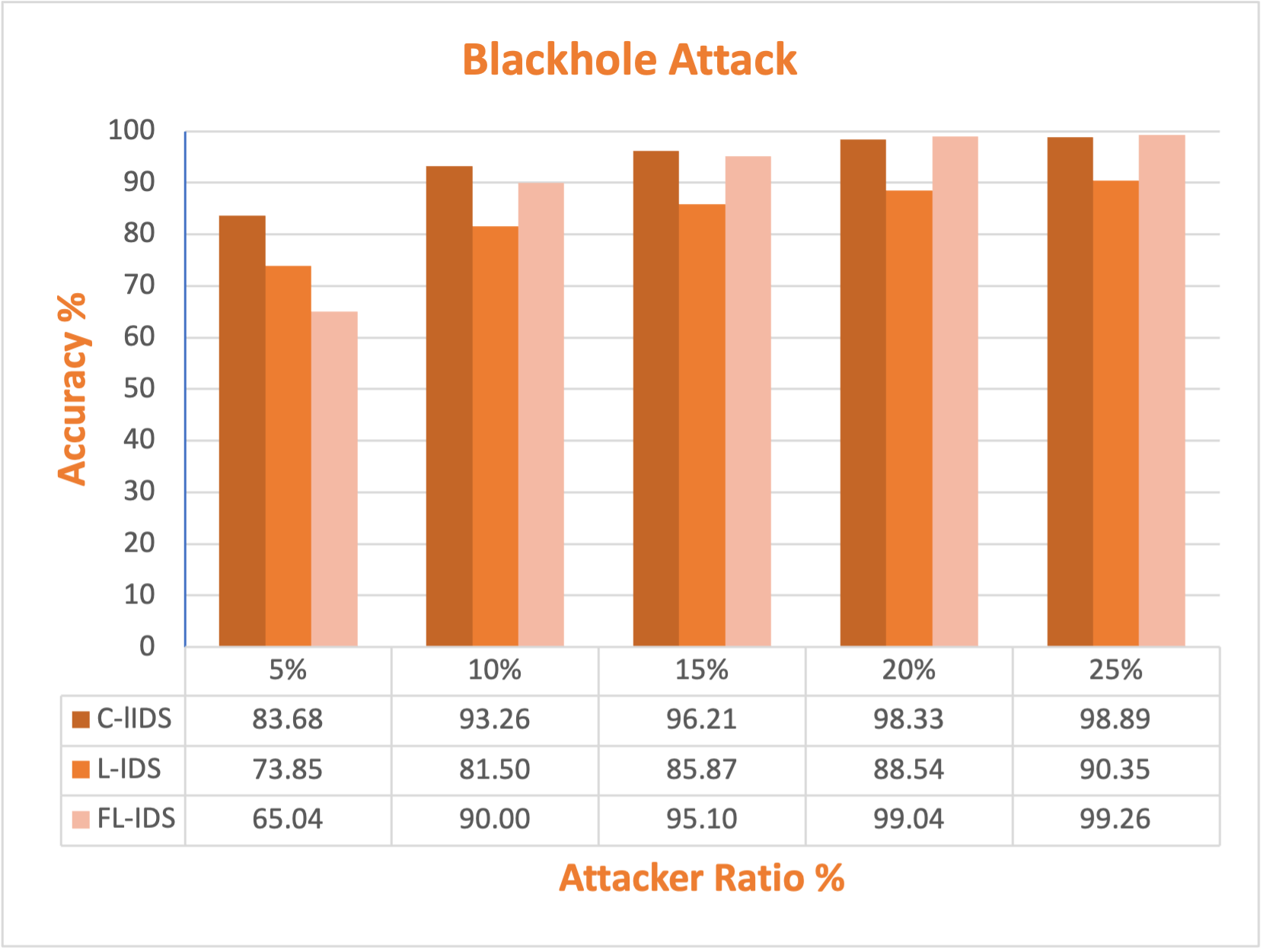}
    \caption{Blackhole Attack}
    \label{fig:blackhole} % Properly associated with the subcaption
  \end{subfigure}
  
  % Spacing between rows
  \vspace{1em}
  
  % Second row: Flooding image centered below
  \begin{subfigure}[b]{0.5\linewidth}
    \centering
    \includegraphics[width=.9\linewidth]{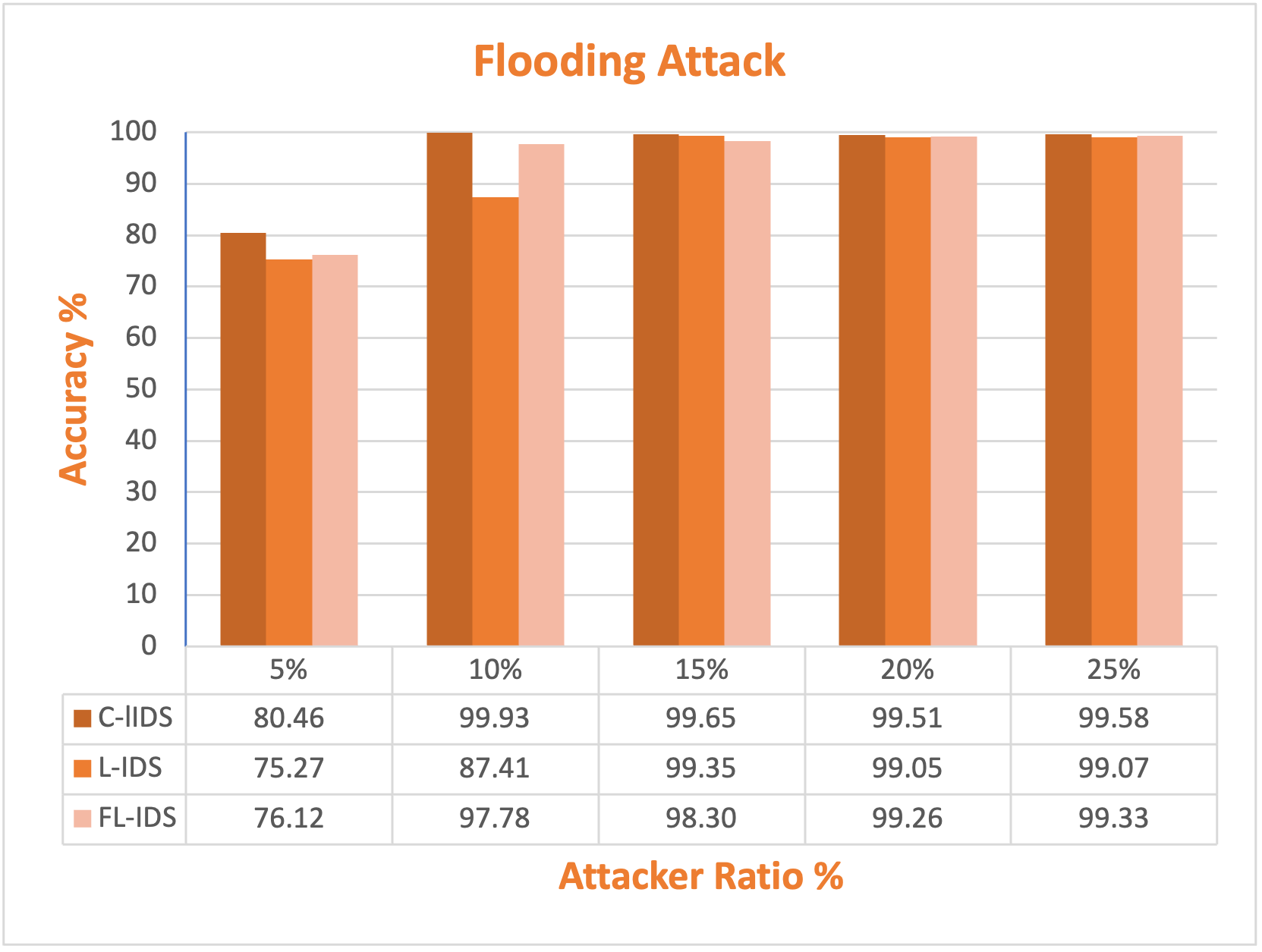}
    \caption{Flooding Attack}
    \label{fig:flooding} % Properly associated with the subcaption
  \end{subfigure}

  \caption{Comparison of C-IDS, L-IDS, and FL-IDS in Detecting Attacks. Subfigures (a), (b), and (c) show individual attack types.}
  \label{fig:fig4} % Label for the entire figure
\end{figure*}

Three IDSs (C-IDS, L-IDS, and FL-IDS) are compared in Figure \ref{fig:sinkhole}. With a low attacker ratio of 5\%, all IDSs perform poorly. However, their performance improves significantly as the attacker ratio increases. Although there is a substantial gap between C-IDS (82.85\%) and the others (L-IDS: 72. 99\%, FL-IDS: 70. 44\%) in the attacker ratio of 5\%, this gap narrows as the attacker ratio in the network increases. Especially at an attacker ratio of 15 or higher, the performance gap between C-IDS and FL-IDS narrows to less than 1.5\%. \myrev{ In federated learning, a higher proportion of attacker nodes exposes individual nodes to a wider range of anomalous data and attack patterns, increasing their exposure to attack behaviors. This broader exposure improves the detection accuracy at individual nodes, allowing them to learn and adapt to a more diverse set of attack characteristics. Consequently, the global model, which aggregates these enhanced local models, demonstrates improved overall effectiveness in detecting attacks, thereby reducing the performance difference between FL-IDS and C-IDS as attacker ratios increase.} On the other hand, L-IDS consistently performs much worse than the others, even achieves 90.27\% accuracy at the highest attacker ratio, while FL-IDS and C-IDS obtain 97. 70\% and 99. 03\% accuracy, respectively. While some local monitoring nodes such as close to the sinkhole in the network might be more affected from the attack, other nodes can be less affected, or even isolated nodes might not be affected at all, hence this situation is reflected in the local IDS' results.

\begin{figure*}[h]

  \begin{minipage}{\linewidth}
  \includegraphics[width=.5\linewidth]{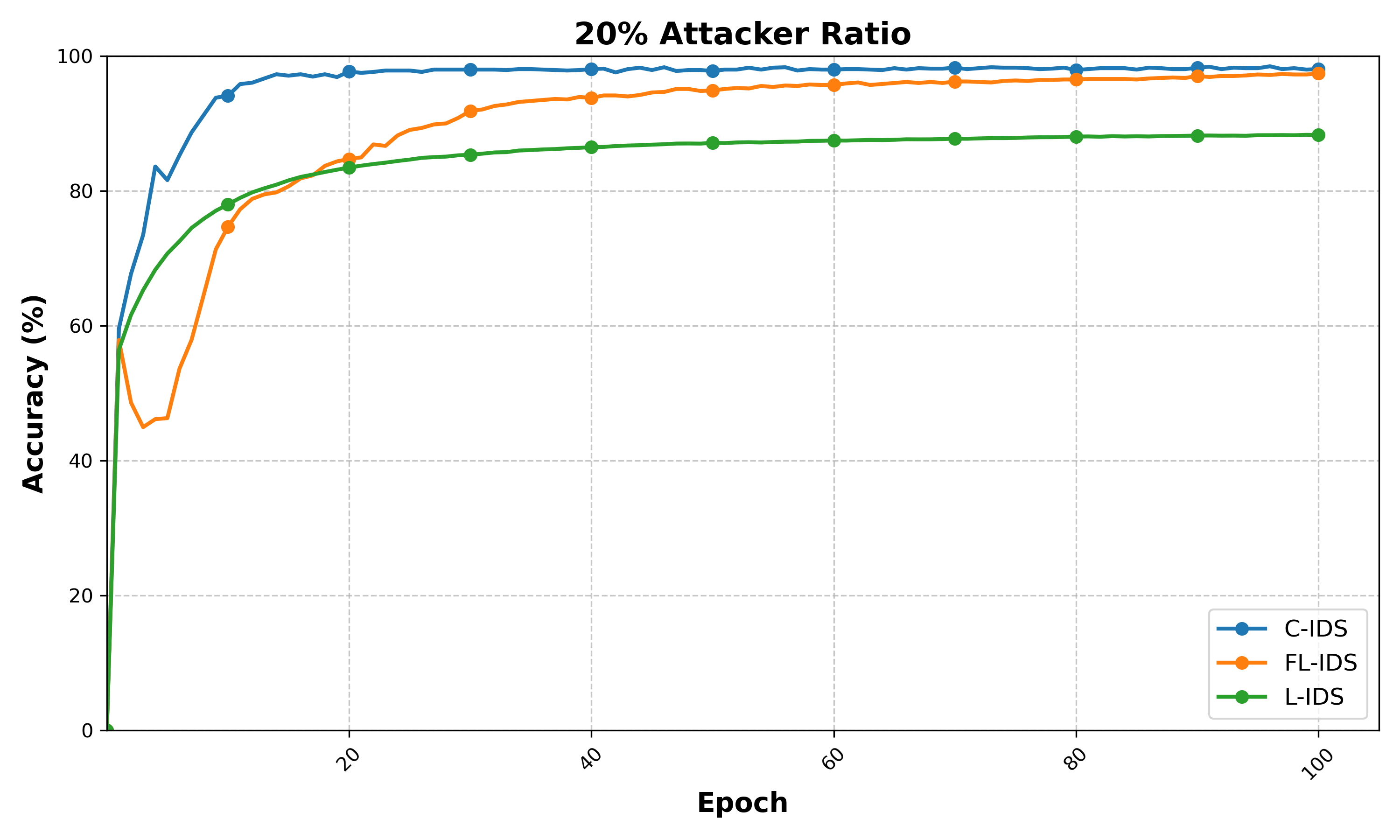}\hfill
  \includegraphics[width=.5\linewidth]{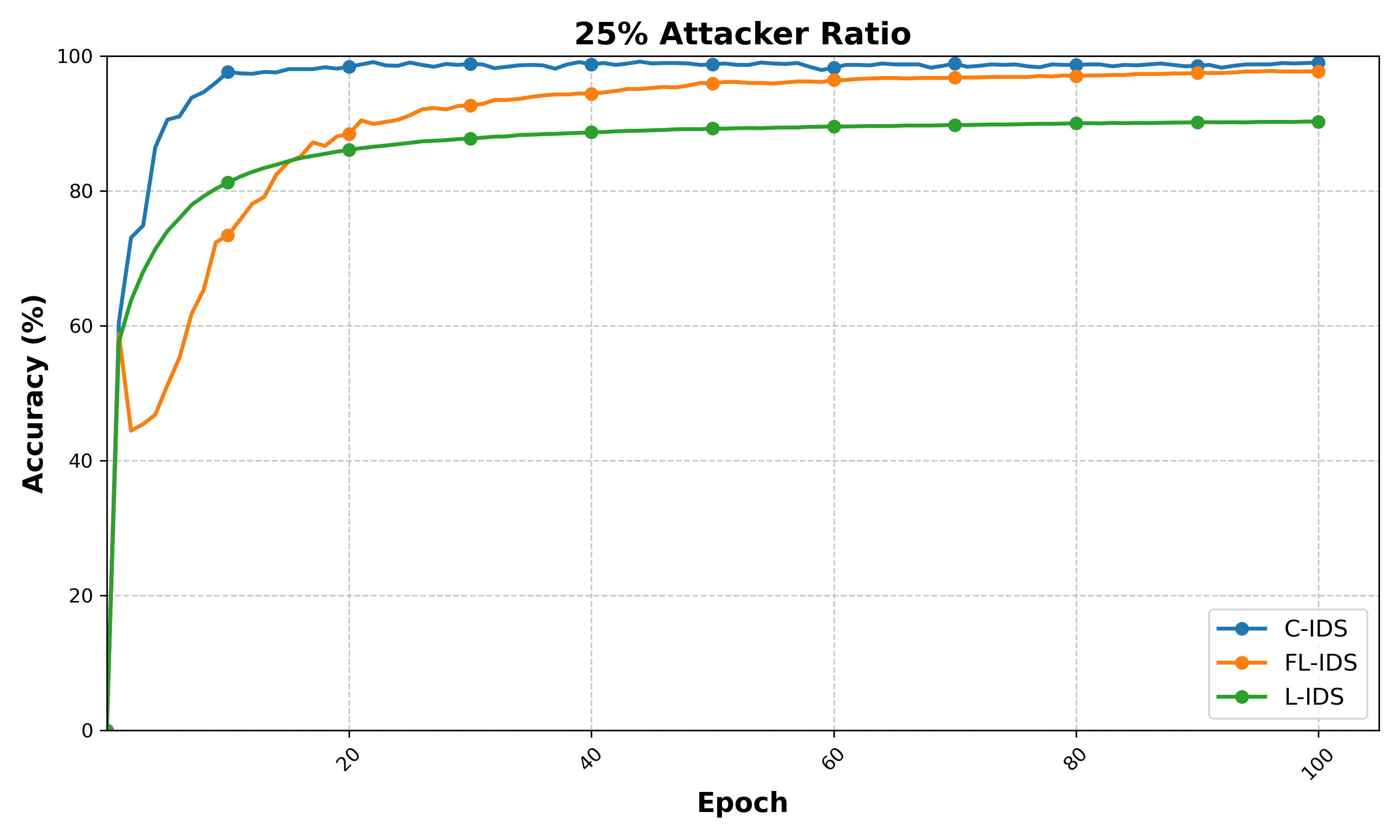}\hfill
  \end{minipage}%
  
  \caption{Convergence of C-IDS, L-IDS and FL-IDS in Detecting Sinkhole Attack for 20\% and 25\% Attacker Ratios}
  \label{fig:fig5}
\end{figure*}

These results highlight the central model's advantage in accessing data from all clients, enabling high accuracy in detecting sinkhole attacks. However, as mentioned earlier, the performance of FL-IDS converges to a level close to that of C-IDS, with a difference of up to 1.5\% at higher attacker ratios ($\geq$ 15\%). Figure \ref{fig:fig5} displays the convergence graphs of each model for attacker ratios of 20\% and 25\%. The graphs reveal that FL-IDS converges to Central IDS at approximately the 60th epoch.

We also employed an approach called Bias towards specific clients (BTSC) in order to improve the performance of FL-IDS at low attacker ratios. The approach involves giving more weights to the better detecting clients, as some clients may be better at detecting attacks than others due to the dynamic nature of FANETs. In the federated aggregation, \myrev{20\%} of the clients are taken, allowing us to focus on the best detectors. The results obtained with BTSC are shown and compared with the FL-IDS trained with all clients in Table \ref{Table:7}. As shown in the results, when the federated model is trained with the best clients, FL-IDS outperform C-IDS even at the lowest attacker ratio.

\begin{table}[h]
\centering
\caption{Performance Improvement with BTSC in Detecting Sinkhole Attack}
\label{Table:7}
\resizebox{0.5\columnwidth}{!}{%
\begin{tabular}{|c|c|c|}
\hline
Attacker   Ratio (\%) & FL-IDS (without BTSC) & FL-IDS (BTSC) \\ 
\hline
5 & 70.44\% & 82.96\% \\ 
\hline
10 & 89.18\% & 95.26\% \\ 
\hline
15 & 94.07\% & 97.63\% \\ 
\hline
20 & 97.41\% & 99.48\% \\ 
\hline
25 & 97.70\% & 99.62\% \\
\hline
\end{tabular}%
}
\end{table}
\subsubsection{Blackhole Attack}

All results for the three IDSs on the detection of blackhole attacks are presented in Figure \ref{fig:blackhole}. With a low attacker ratio of 5\%, all IDSs exhibit inadequate performance in detecting blackhole attacks. The federated learning-based approach, in particular, performs the worst with a success rate of only 65.03\% at this attacker ratio. FL-IDS, being based on decentralized data sources, can only share their weights with the global server. Since attacker nodes cannot influence all nodes uniformly across the network, the data distribution is uneven, leading to lower accuracy at lower attacker ratios. The best performance at the lowest attacker ratio is achieved by C-IDS, reaching 83.68\%. C-IDS benefits from its access to all network-generated data, providing a comprehensive view of network traffic and patterns, which contributes to its high accuracy in detecting attacks.

The performance gap between FL-IDS and C-IDS decreases as the attacker ratio increases in the network. Specifically, when the attacker ratio reached 25\%, the difference in accuracy between C-IDS and FL-IDS reduced to less than 0.5\%. In contrast, L-IDS consistently underperformed the other two. The key distinction is that Local IDSs exclusively process their own data and lack a framework for inter-client communication. This lack of communication capabilities accounts for their comparatively lower accuracy.

However, with an increase in the attacker ratio, accuracy tends to improve across all approaches. This is attributed to a higher number of attacker nodes in the network, introducing more anomaly data and patterns for deep learning algorithms to analyze. Consequently, all three approaches exhibit increased performance as the attacker ratio increases. As a result, the best performance for L-IDS (90. 35\%) is achieved with the highest attacker ratio, as expected.

\begin{figure*}[!t]

  \begin{minipage}{\linewidth}
  \includegraphics[width=.5\linewidth]{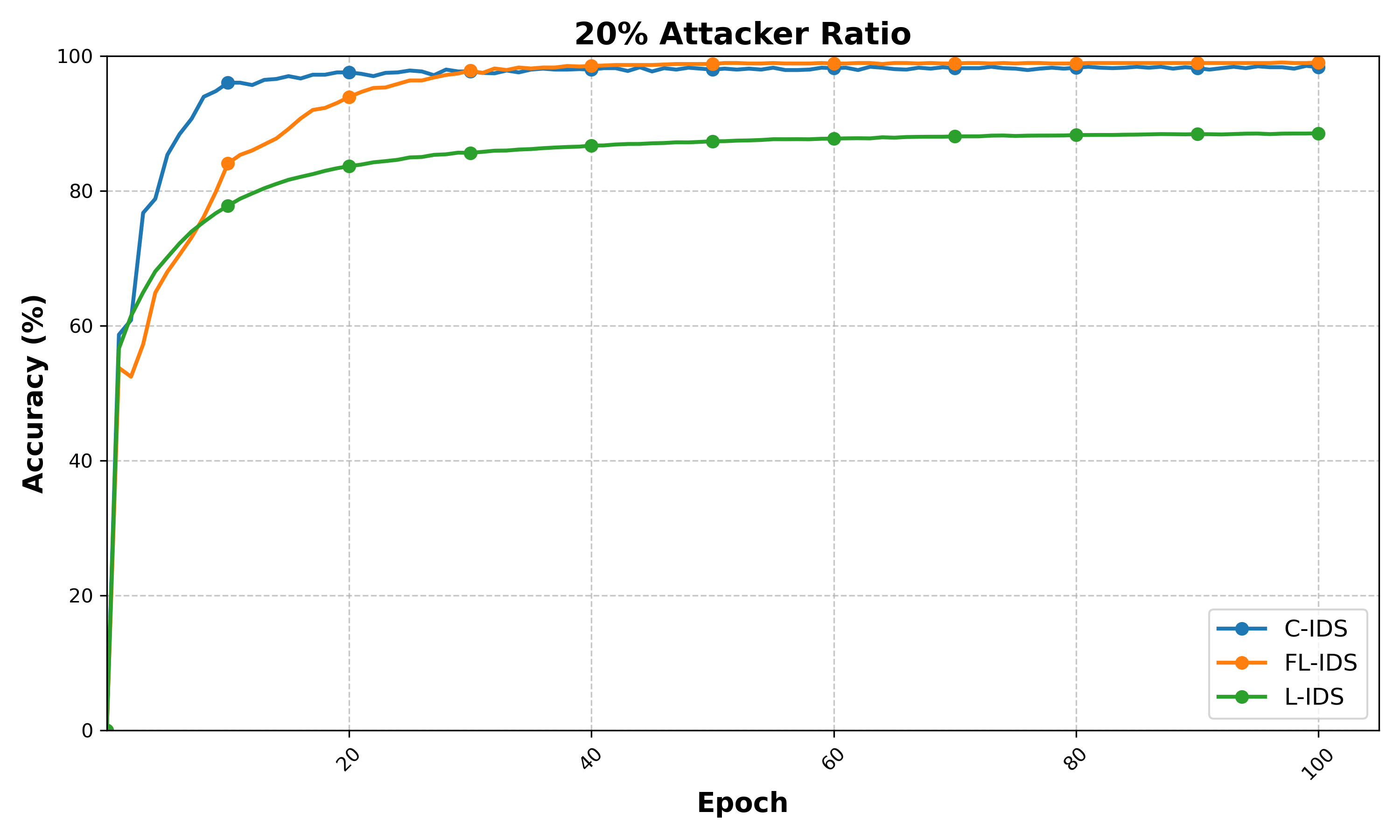}\hfill
  \includegraphics[width=.5\linewidth]{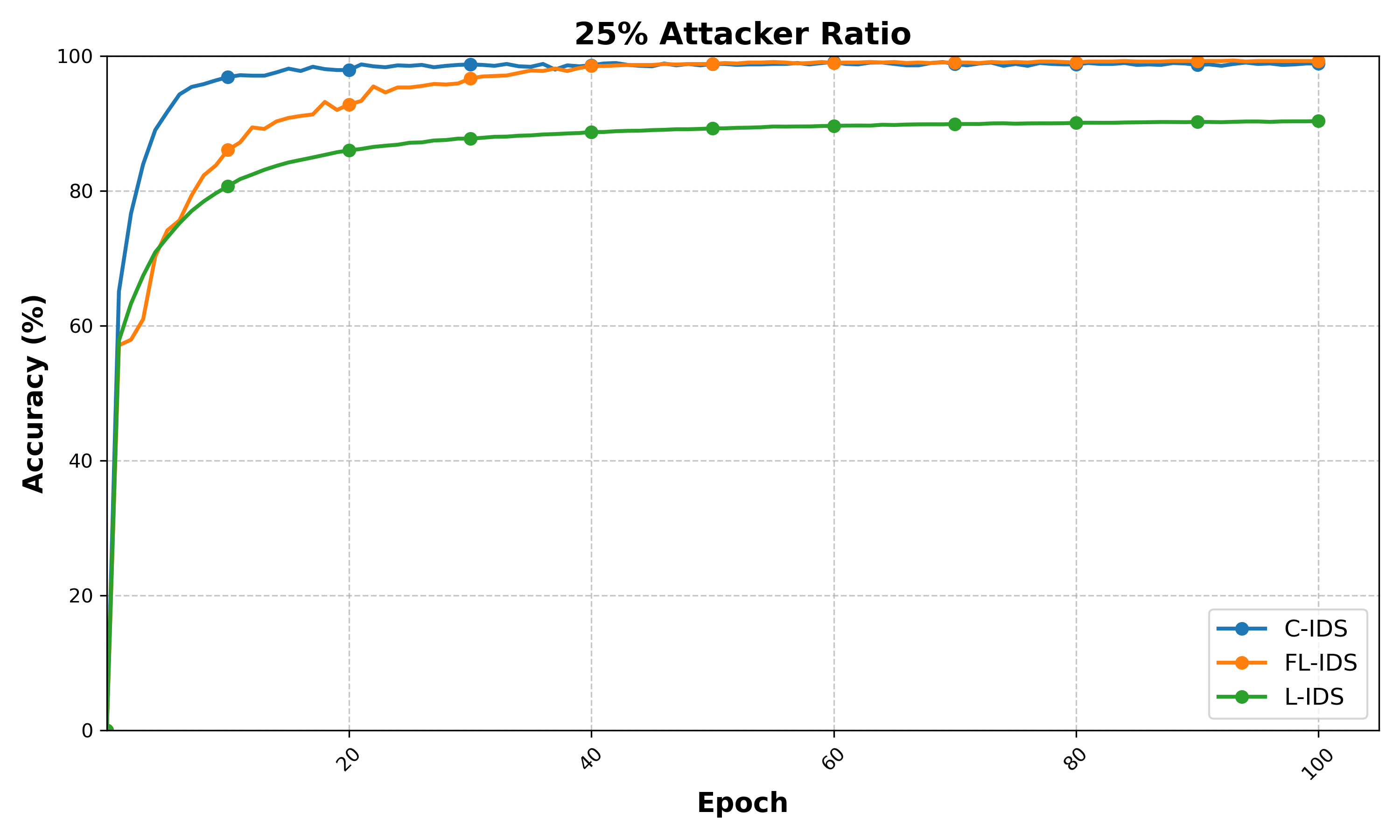}\hfill
  \end{minipage}%
  
  \caption{Convergence of C-IDS, L-IDS and FL-IDS in Detecting Blackhole Attack for 20\% and 25\% Attacker Ratios}
  \label{fig:fig3}
\end{figure*}

In Figure \ref{fig:fig3}, the accuracy of the IDSs is depicted for each epoch, revealing that FL-IDS closely approached the accuracy of C-IDS at different epochs for various attacker ratios. In particular, FL-IDS reached a level of accuracy comparable to C-IDS at approximately the 30th epoch for the 20\% and 25\% attacker ratios. However, for the 5\% attacker ratio, FL-IDS cannot match C-IDS's accuracy at any epoch, with C-IDS consistently outperforming FL-IDS in subsequent epochs. Furthermore, it should be noted that L-IDS consistently yielded lower results compared to both types of IDS.

The performance of FL-IDS is improved by employing the BTSC method as shown in Table \ref{Table:8}. When FL-IDS is trained with the weight of the best clients, it achieves high accuracy results, surpassing C-IDS. On the other hand, while its performance reaches 80\% accuracy at the lowest ratio, C-IDS still achieves the best performance.

\begin{table}[!th]
\centering
\caption{Performance Improvement with BTSC in Detecting Blackhole Attack}
\label{Table:8}
\resizebox{0.5\columnwidth}{!}{%
\begin{tabular}{|c|c|c|}
\hline
Attacker   Ratio (\%) & FL-IDS (without BTSC) & FL-IDS (BTSC) \\ \hline
5 & 65.04\% & 79.80\% \\ \hline
10 & 90.00\% & 95.70\% \\ \hline
15 & 95.10\% & 98.53\% \\ \hline
20 & 99.04\% & 99.56\% \\ \hline
25 & 99.26\% & 99.85\% \\ \hline
\end{tabular}%
}
\end{table}

\subsubsection{Flooding Attack}

Figure \ref{fig:flooding} shows the comparative result of three IDS  approaches for detecting flooding attack. Differently from other attacks IDSs based on central or federated learning-based approaches obtain high accuracy even when the attacker ratio is as low as 10\%. C-IDS (98. 13\%) and FL-IDS (97. 78\%) have a high performance since many nodes are affected by the attack and participate in training. On the other hand, L-IDS (87.41\%) show considerably lower performance than C-IDS and FL-IDS in the 10\% of the attacker ratio. This can be a result of some nodes which might be less affected by the attack than others due to their positions. For instance, some nodes might receive fewer flooding packets due to being more isolated and having a few neighbour nodes only. However, when the attacker ratio is 15\% or higher, L-IDS almost reaches the results of C-IDS and FL-IDS, as the number of nodes affected by the attack also increases.

\begin{figure*}[!t]
  \begin{minipage}{\linewidth}
  \includegraphics[width=.5\linewidth]{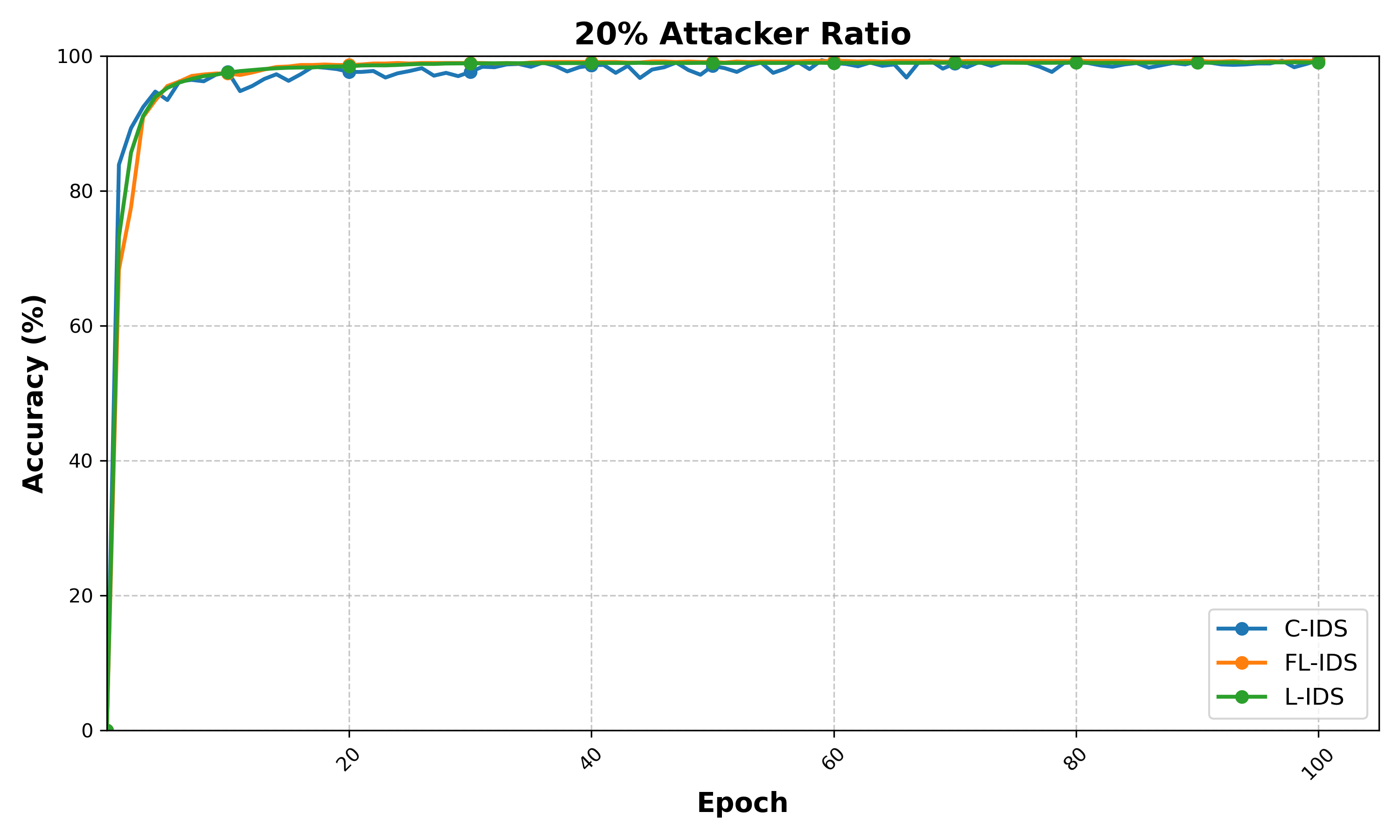}\hfill
  \includegraphics[width=.5\linewidth]{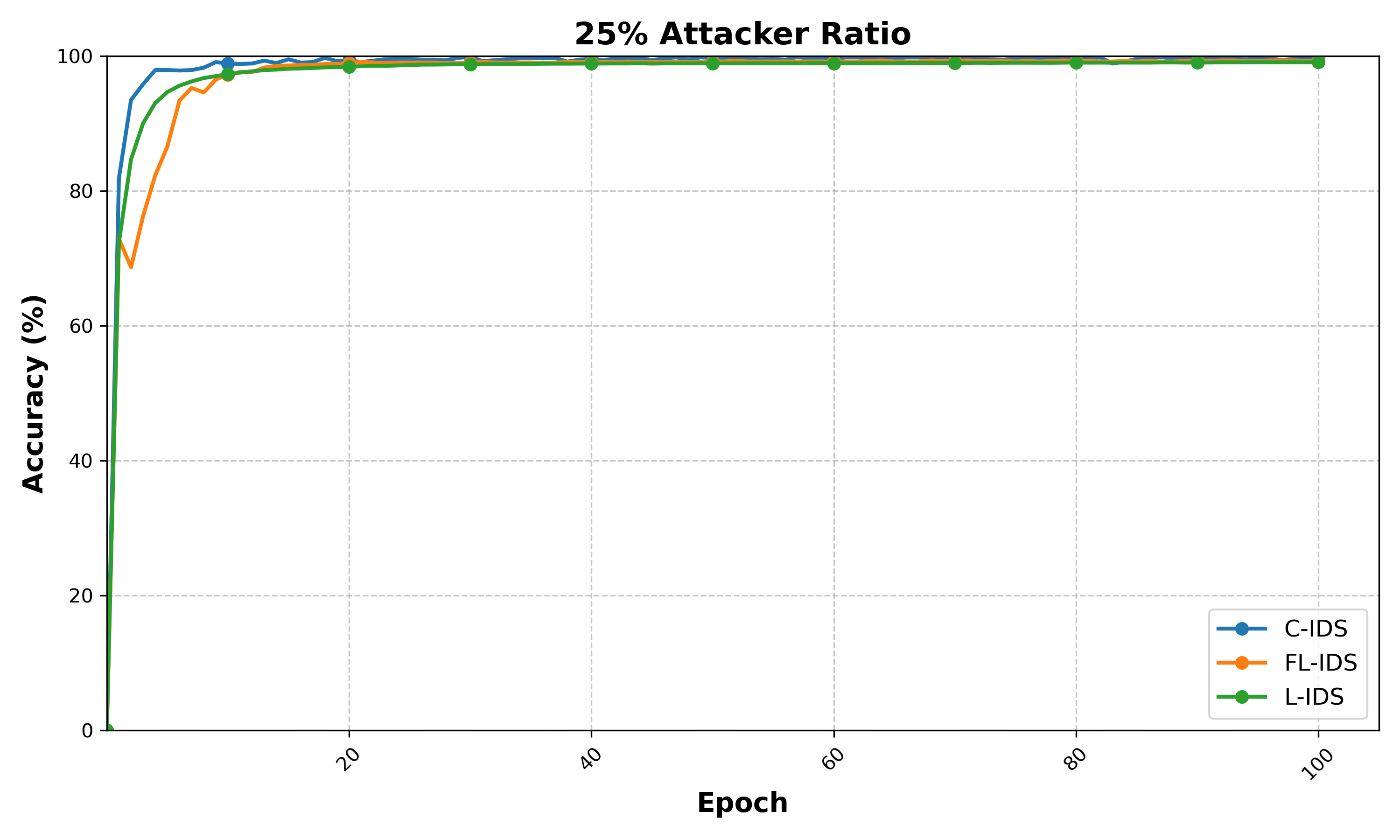}\hfill
  \end{minipage}%
  \caption{Convergence of C-IDS, L-IDS and FL-IDS in Detecting Flooding Attack for 20\% and 25\% Attacker Ratios}
  \label{fig:fig7}
\end{figure*}

The results reveal that C-IDS benefits from having access to all client data, enabling it to achieve high accuracy, even at lower attacker ratios. In comparison to other attack scenarios, L-IDS maintains competitive performance to C-IDS and FL-IDS, considering the characteristics of flooding attacks that can affect most of the nodes in the network. Its performance is consistent, and its accuracy values are close to those of other IDSs. Figure \ref{fig:fig7} shows that L-IDS converges to C-IDS and FL-IDS at early epochs (approximately at the 6th epoch for 20\% attacker ratio and at the 11th epoch for 25\% attacker ratio). The obtained results demonstrate that all three IDSs show high results and reduce the performance gap among them at  earlier epochs.

Finally, although the accuracy results are already high for flooding attacks, FL-IDS has been further improved with the BTSC method, as indicated in Table \ref{Table:10}.

\begin{table}[!h]
\centering
\small
\caption{Performance Improvement with BTSC in Detecting Flooding Attack}
\label{Table:10}
\resizebox{.5\columnwidth}{!}{%
\begin{tabular}{|c|c|c|}
\hline
Attacker   Ratio (\%) & FL-IDS (without BTSC) & FL-IDS (BTSC) \\ \hline
5 & 76.12\% & 78.57\% \\ \hline
10 & 97.78\% & 99.11\% \\ \hline
15 & 98.30\% & 99.26\% \\ \hline
20 & 99.26\% & 99.55\% \\ \hline
25 & 99.33\% & 99.63\% \\ \hline
\end{tabular}%
}

\end{table}

\subsubsection{Detection Rate vs. False Positive Rate}

Table \ref{tab:DrFpr} provides information on the detection rate (DR) and the false positive rate (FPR) for each type of attack. The results reveal that all IDS models exhibit relatively high false positive rates at the lowest attacker ratio. However, as the attacker ratio increases, the FPR tends to decrease to reasonable levels. In particular, when the network contains a significant number of attackers (around 20\% to 25\%), each IDS model's performance approaches an ideal scenario. In flooding attacks, IDS converge to the ideal scenario even when the attacker ratio is as low as 10\% on the network, since this attack causes significant damage to the network. Unlike other attacks, flooding attacks affect almost all nodes, making it easier to differentiate abnormal data from normal data. It is worth highlighting that, on average, blackhole attacks perform slightly better than sinkhole attacks. While these attacks share similarities in directing traffic through attacker nodes, blackhole attacks also drop the packets they attract. This action might lead to an interruption in network traffic, as legitimate packets fail to reach their intended destinations. This disruption in network communications is more readily detectable because it directly impacts connectivity and traffic flow.

\begin{table*}[]
\centering
\caption{Evaluation Results of C-IDS, L-IDS and FL-IDS in terms of DR and FPR}
\label{tab:DrFpr}
\resizebox{.5\columnwidth}{!}{%
\begin{tabular}{|c|c|rr|rr|rr|}
\hline
\multicolumn{1}{|l|}{} & \multicolumn{1}{l|}{} & \multicolumn{2}{c|}{C-IDS} & \multicolumn{2}{c|}{L-IDS} & \multicolumn{2}{c|}{FL-IDS} \\ \hline
\multicolumn{1}{|l|}{Attack Type} & \multicolumn{1}{l|}{Attack ratio (\%)} & \multicolumn{1}{c|}{DR} & \multicolumn{1}{c|}{FPR} & \multicolumn{1}{c|}{DR} & \multicolumn{1}{c|}{FPR} & \multicolumn{1}{c|}{DR} & \multicolumn{1}{c|}{FPR} \\ \hline
\multirow{5}{*}{Sinkhole} & 5 & \multicolumn{1}{r|}{81.20\%} & 18.80\% & \multicolumn{1}{r|}{75.45\%} & 24.55\% & \multicolumn{1}{r|}{83.00\%} & 17.00\% \\ \cline{2-8} 
 & 10 & \multicolumn{1}{r|}{93.97\%} & 6.03\% & \multicolumn{1}{r|}{92.23\%} & 7.77\% & \multicolumn{1}{r|}{95.67\%} & 4.33\% \\ \cline{2-8} 
 & 15 & \multicolumn{1}{r|}{94.27\%} & 5.73\% & \multicolumn{1}{r|}{95.27\%} & 4.73\% & \multicolumn{1}{r|}{98.01\%} & 1.99\% \\ \cline{2-8} 
 & 20 & \multicolumn{1}{r|}{98.28\%} & 1.72\% & \multicolumn{1}{r|}{97.65\%} & 2.35\% & \multicolumn{1}{r|}{99.31\%} & 0.69\% \\ \cline{2-8} 
 & 25 & \multicolumn{1}{r|}{98.56\%} & 1.44\% & \multicolumn{1}{r|}{99.02\%} & 0.98\% & \multicolumn{1}{r|}{99.48\%} & 0.52\% \\ \hline
\multirow{5}{*}{Blackhole} & 5 & \multicolumn{1}{r|}{83.73\%} & 16.27\% & \multicolumn{1}{r|}{68.29\%} & 31.71\% & \multicolumn{1}{r|}{75.17\%} & 24.83\% \\ \cline{2-8} 
 & 10 & \multicolumn{1}{r|}{93.07\%} & 6.93\% & \multicolumn{1}{r|}{87.45\%} & 12.55\% & \multicolumn{1}{r|}{93.10\%} & 6.90\% \\ \cline{2-8} 
 & 15 & \multicolumn{1}{r|}{95.58\%} & 4.42\% & \multicolumn{1}{r|}{93.72\%} & 6.28\% & \multicolumn{1}{r|}{95.16\%} & 4.84\% \\ \cline{2-8} 
 & 20 & \multicolumn{1}{r|}{97.87\%} & 2.13\% & \multicolumn{1}{r|}{96.18\%} & 3.82\% & \multicolumn{1}{r|}{99.48\%} & 0.52\% \\ \cline{2-8} 
 & 25 & \multicolumn{1}{r|}{99.20\%} & 0.80\% & \multicolumn{1}{r|}{97.38\%} & 2.62\% & \multicolumn{1}{r|}{99.66\%} & 0.34\% \\ \hline
\multirow{5}{*}{Flooding} & 5 & \multicolumn{1}{r|}{70.89\%} & 29.11\% & \multicolumn{1}{r|}{61.44\%} & 38.56\% & \multicolumn{1}{r|}{63.44\%} & 36.56\% \\ \cline{2-8} 
 & 10 & \multicolumn{1}{r|}{98.56\%} & 1.44\% & \multicolumn{1}{r|}{98.18\%} & 1.82\% & \multicolumn{1}{r|}{98.14\%} & 1.86\% \\ \cline{2-8} 
 & 15 & \multicolumn{1}{r|}{99.29\%} & 0.71\% & \multicolumn{1}{r|}{98.58\%} & 1.42\% & \multicolumn{1}{r|}{98.79\%} & 1.21\% \\ \cline{2-8} 
 & 20 & \multicolumn{1}{r|}{99.42\%} & 0.58\% & \multicolumn{1}{r|}{99.03\%} & 0.97\% & \multicolumn{1}{r|}{99.48\%} & 0.52\% \\ \cline{2-8} 
 & 25 & \multicolumn{1}{r|}{99.86\%} & 0.14\% & \multicolumn{1}{r|}{99.61\%} & 0.39\% & \multicolumn{1}{r|}{99.61\%} & 0.39\% \\ \hline
\end{tabular}}
\end{table*}

\section{Discussion and Limitations} \label{discusandlimitation}

In this section, we evaluate the proposed FL-IDS in terms of its effectiveness, efficiency, communication cost, response time, privacy and security. We also discuss the comparison with C-IDS and L-IDS based on these criteria.

\subsection{Accuracy}
The results discussed above clearly show that when we centrally obtain local data and train a model using all these data, we can better differentiate attacks from benign network traffic. Therefore,  Central IDS (C-IDS) shows the best performance for each type of attack. However, Federated IDS (FL-IDS) achieves competitive results with C-IDS. In particular, at higher attacker ratios, FL-IDS can approach the accuracy of C-IDS \myrev{because the increased number of attacker nodes introduces a richer set of anomalous patterns across the network. This leads to more robust local models being trained in the data set, which are then aggregated into a global model that better generalizes attack detection across the entire network.} When the attacker ratio increases, more devices affected by such attacks can positively participate in FL-IDS training. This is also supported by the improvement obtained with the use of BTSC, which selectively considers the best clients for training FL-IDS.

Local IDSs (L-IDS) have a lower success rate on average compared to other IDSs. The performance of L-IDS is calculated as the average of each local client. Therefore, L-IDS only shows results comparable to those of C-IDS and FL-IDS in detecting flooding attacks due to the nature of this attack type, which affects most nodes in the network.  However, when the impact of an attack is limited, as is the case with sinkhole and blackhole attacks that affect specific victims and their surroundings, the average accuracy of local IDSs may not be high. This discrepancy arises because, while some clients can effectively detect attacks due to their location (such as being situated in active routes or denser areas), others may have less success.

The performance of IDS on the nodes has been compared in Tables \ref{Table:12}, \ref{tab:blackholeLocalClients}, and \ref{Table:13} for sinkhole, blackhole, and flooding attacks, respectively\myrev{, based on accuracy}. These tables present the performance of the best and worst IDSs for both L-IDS and FL-IDS \myrev{when deployed on different nodes}, along with their average performance. \myrev{In traditional L-IDS, where each node operates solely on its local data, we observe a significant disparity between the worst and best node performance of detection. This variation underscores the limitations of L-IDS, particularly in scenarios where nodes have insufficient data or lack diversity in their training sets.} \myrev{FL-IDS outperforms L-IDS in terms of the worst-performing node across all attacker ratios and attack types. However, the best-performing node in L-IDS achieves a higher detection accuracy than the best-performing node in FL-IDS. This difference can be attributed to the fact that in L-IDS, each node is trained exclusively on its own local data. As a result, when the test data closely matches the node’s local attack patterns, L-IDS can achieve significantly higher detection performance. However, this advantage is node specific and highly dependent on the distribution and density of local attack data.}

\myrev{  One notable observation is that under blackhole and sinkhole attacks, when the attacker ratio exceeds 10\%, worst-performing node in FL-IDS that receive feedback from the global model as part of federated learning demonstrate better performance than the average accuracy of L-IDS. } However, when dealing with flooding attacks and as the number of affected nodes increases, L-IDS outperforms FL-IDS, even at lower attacker ratios. \myrev{ To further enhance the performance of FL-IDS, it is recommended to use BTSC for training IDS in federated learning, particilarly in scenerios where a significant portion of the network's clients remain unaffected by the attack.}

The use of BTSC results in improved performance, particularly when a significant portion of the network's clients remain unaffected by the attack. \myrev{By prioritizing high-performing clients based on their detection accuracy, BTSC ensures that only the most relevant and generalizable model updates are incorporated during the federated aggregation process. This leads to significant improvements in detection accuracy. In addition, BTSC reduces communication overhead by limiting the number of contributing clients, which is particularly beneficial in resource-constrained FANET environments. Additionally, the approach helps lower energy consumption by minimizing the need for extensive data transmission. The overall performance improvements, along with the robust generalization demonstrated through validation on a diverse test dataset, underscore the validity of the proposed approach in improving the efficiency and effectiveness of FL-IDS in dynamic and heterogeneous FANETs.}

\color{black}

\begin{table}[th!]
\centering
\caption{\myrev{Best/Worst Accuracy of IDSs in Detecting in Sinkhole Attacks}}
\label{Table:12}
\resizebox{.5\columnwidth}{!}{%
\begin{tabular}{|c|cc|cc|c|c|}
\hline
\multirow{2}{*}{Attacker   Ratio (\%)} & \multicolumn{2}{c|}{L-IDS} & \multicolumn{2}{c|}{FL-IDS} & L-IDS & FL-IDS \\ \cline{2-7} 
 & \multicolumn{1}{l|}{Worst} & Best & \multicolumn{1}{l|}{Worst} & Best & Average & Average \\ \hline
5 & \multicolumn{1}{l|}{58.61\%} & 98.33\% & \multicolumn{1}{l|}{65.08\%} & 80.63\% & 72.99\% & 70.68\% \\ \hline
10 & \multicolumn{1}{l|}{72.50\%} & 98.96\% & \multicolumn{1}{l|}{80.18\%} & 95.23\% & 81.19\% & 88.12\% \\ \hline
15 & \multicolumn{1}{l|}{76.18\%} & 99.51\% & \multicolumn{1}{l|}{86.93\%} & 97.74\% & 85.40\% & 93.53\% \\ \hline
20 & \multicolumn{1}{l|}{83.06\%} & 99.93\% & \multicolumn{1}{l|}{90.04\%} & 98.92\% & 88.28\% & 96.43\% \\ \hline
25 & \multicolumn{1}{l|}{88.54\%} & 99.93\% & \multicolumn{1}{l|}{91.79\%} & 99.41\% & 90.27\% & 97.36\% \\ \hline
\end{tabular}%
}

\end{table}

\begin{table}[th!]
\centering
\caption{\myrev{Best/Worst Accuracy of IDSs in Detecting in Blackhole Attacks}}
\label{tab:blackholeLocalClients}
\resizebox{.5\columnwidth}{!}{%
\begin{tabular}{|c|cc|cc|c|c|}
\hline
\multirow{2}{*}{Attacker   Ratio (\%)} & \multicolumn{2}{c|}{L-IDS} & \multicolumn{2}{c|}{FL-IDS} & L-IDS & FL-IDS \\ \cline{2-7} 
 & \multicolumn{1}{l|}{Worst} & Best & \multicolumn{1}{l|}{Worst} & Best & Average & Average \\ \hline
5 & \multicolumn{1}{l|}{58.40\%} & 97.92\% & \multicolumn{1}{l|}{61.38\%} & 74.20\% & 73.85\% & 66.08\% \\ \hline
10 & \multicolumn{1}{l|}{71.53\%} & 99.10\% & \multicolumn{1}{l|}{80.69\%} & 95.30\% & 81.50\% & 88.94\% \\ \hline
15 & \multicolumn{1}{l|}{85.82\%} & 99.44\% & \multicolumn{1}{l|}{86.78\%} & 98.41\% & 85.87\% & 94.33\% \\ \hline
20 & \multicolumn{1}{l|}{87.57\%} & 99.65\% & \multicolumn{1}{l|}{92.23\%} & 99.34\% & 88.54\% & 97.49\% \\ \hline
25 & \multicolumn{1}{l|}{88.33\%} & 99.86\% & \multicolumn{1}{l|}{94.32\%} & 99.53\% & 90.35\% & 98.13\% \\ \hline
\end{tabular}%
}

\end{table}

\begin{table}[th!]
\centering
\caption{\myrev{Best/Worst Accuracy of IDSs in Detecting in Flooding Attacks}}
\label{Table:13}
\resizebox{.5\columnwidth}{!}{%
\begin{tabular}{|c|cc|cc|c|c|}
\hline
\multirow{2}{*}{Attacker   Ratio (\%)} & \multicolumn{2}{c|}{L-IDS} & \multicolumn{2}{c|}{FL-IDS} & L-IDS & FL-IDS \\ \cline{2-7} 
 & \multicolumn{1}{l|}{Worst} & Best & \multicolumn{1}{l|}{Worst} & Best & Average & Average \\ \hline
5 & \multicolumn{1}{l|}{70.49\%} & 85.39\% & \multicolumn{1}{l|}{72.21\%} & 81.48\% & 75.27\% & 77.44\% \\ \hline
10 & \multicolumn{1}{l|}{89.58\%} & 99.79\% & \multicolumn{1}{l|}{96.41\%} & 98.30\% & 87.41\% & 99.40\% \\ \hline
15 & \multicolumn{1}{l|}{94.03\%} & 99.93\% & \multicolumn{1}{l|}{98.73\%} & 99.86\% & 99.35\% & 99.39\% \\ \hline
20 & \multicolumn{1}{l|}{94.24\%} & 100\% & \multicolumn{1}{l|}{98.56\%} & 99.79\% & 99.05\% & 98.42\% \\ \hline
25 & \multicolumn{1}{l|}{99.72\%} & 100\% & \multicolumn{1}{l|}{98.37\%} & 99.93\% & 99.07\% & 98.424\% \\ \hline
\end{tabular}%
}

\end{table}

\myrevtwo{We also evaluate the performance of three widely used aggregation algorithms, FedAvg, FedProx \cite{li2020federated}, and FedSGD \cite{li2020preserving} . Our evaluation shows that FedAvg consistently outperforms FedProx across key metrics, including accuracy, DR, and FPR. FedProx is designed to handle environments with significant systems heterogeneity and statistical heterogeneity \cite{li2020federated}. In our setting, however, FedProx's regularization mechanism appears to introduce additional constraints that slow convergence and diminish attack detection performance, leading to worse results compared to FedAvg. FedSGD exhibits distinct performance characteristics compared to both FedAvg and FedProx. Unlike these methods, FedSGD updates the global model after each mini-batch, resulting in increased communication frequency but reduced local computation. The results show that FedSGD maintains an exceptionally low FPR; however, its DR is significantly lower than both FedAvg and FedProx, indicating weaker performance in identifying attacks. This reduced DR may stem from the lack of local model training, limiting the model's ability to learn robust attack patterns.}

\begin{table}[]
\centering
\caption{\myrevtwo{Comparative Evaluation of Aggregation Algorithms}}
\label{tab:FedAlgr}
\resizebox{.8\columnwidth}{!}{%
% \usepackage{multirow}

\begin{tabular}{cclrrllllrr}
\hline
\multicolumn{1}{|l|}{} &
  \multicolumn{1}{l|}{} &
  \multicolumn{3}{c|}{FedAvg} &
  \multicolumn{3}{c|}{FedProx} &
  \multicolumn{3}{c|}{FedSGD} \\ \hline
\multicolumn{1}{|l|}{Attack Type} &
  \multicolumn{1}{l|}{Attack ratio (\%)} &
  \multicolumn{1}{l|}{Accuracy} &
  \multicolumn{1}{c|}{DR} &
  \multicolumn{1}{c|}{FPR} &
  \multicolumn{1}{l|}{Accuracy} &
  \multicolumn{1}{c|}{DR} &
  \multicolumn{1}{c|}{FPR} &
  \multicolumn{1}{l|}{Accuracy} &
  \multicolumn{1}{c|}{DR} &
  \multicolumn{1}{c|}{FPR} \\ \hline
\multicolumn{1}{|c|}{\multirow{5}{*}{Sinkhole}} &
  \multicolumn{1}{c|}{5} &
  \multicolumn{1}{l|}{70.44\%} &
  \multicolumn{1}{r|}{83.00\%} &
  \multicolumn{1}{r|}{17.00\%} &
  \multicolumn{1}{l|}{56.96\%} &
  \multicolumn{1}{l|}{69.00\%} &
  \multicolumn{1}{l|}{31.00\%} &
  \multicolumn{1}{l|}{56.97\%} &
  \multicolumn{1}{r|}{0.54\%} &
  \multicolumn{1}{r|}{0.45\%} \\ \cline{2-11} 
\multicolumn{1}{|c|}{} &
  \multicolumn{1}{c|}{10} &
  \multicolumn{1}{l|}{89.18\%} &
  \multicolumn{1}{r|}{95.67\%} &
  \multicolumn{1}{r|}{4.33\%} &
  \multicolumn{1}{l|}{55.85\%} &
  \multicolumn{1}{l|}{49.00\%} &
  \multicolumn{1}{l|}{51.00\%} &
  \multicolumn{1}{l|}{66.56\%} &
  \multicolumn{1}{r|}{0.61\%} &
  \multicolumn{1}{r|}{0.38\%} \\ \cline{2-11} 
\multicolumn{1}{|c|}{} &
  \multicolumn{1}{c|}{15} &
  \multicolumn{1}{l|}{94.07\%} &
  \multicolumn{1}{r|}{98.01\%} &
  \multicolumn{1}{r|}{1.99\%} &
  \multicolumn{1}{l|}{65.41\%} &
  \multicolumn{1}{l|}{58.00\%} &
  \multicolumn{1}{l|}{42.00\%} &
  \multicolumn{1}{l|}{71.34\%} &
  \multicolumn{1}{r|}{0.68\%} &
  \multicolumn{1}{r|}{0.31\%} \\ \cline{2-11} 
\multicolumn{1}{|c|}{} &
  \multicolumn{1}{c|}{20} &
  \multicolumn{1}{l|}{97.41\%} &
  \multicolumn{1}{r|}{99.31\%} &
  \multicolumn{1}{r|}{0.69\%} &
  \multicolumn{1}{l|}{65.41\%} &
  \multicolumn{1}{l|}{71.00\%} &
  \multicolumn{1}{l|}{29.00\%} &
  \multicolumn{1}{l|}{72.65\%} &
  \multicolumn{1}{r|}{0.79\%} &
  \multicolumn{1}{r|}{0.20\%} \\ \cline{2-11} 
\multicolumn{1}{|c|}{} &
  \multicolumn{1}{c|}{25} &
  \multicolumn{1}{l|}{97.70\%} &
  \multicolumn{1}{r|}{99.48\%} &
  \multicolumn{1}{r|}{0.52\%} &
  \multicolumn{1}{l|}{66.74\%} &
  \multicolumn{1}{l|}{69.00\%} &
  \multicolumn{1}{l|}{31.00\%} &
  \multicolumn{1}{l|}{74.65\%} &
  \multicolumn{1}{r|}{0.77\%} &
  \multicolumn{1}{r|}{0.22\%} \\ \hline
\multicolumn{1}{|c|}{\multirow{5}{*}{Blackhole}} &
  \multicolumn{1}{c|}{5} &
  \multicolumn{1}{l|}{65.04\%} &
  \multicolumn{1}{r|}{75.17\%} &
  \multicolumn{1}{r|}{24.83\%} &
  \multicolumn{1}{l|}{58.67\%} &
  \multicolumn{1}{l|}{60.00\%} &
  \multicolumn{1}{l|}{40.00\%} &
  \multicolumn{1}{l|}{52.93\%} & 
  \multicolumn{1}{r|}{0.42\%} &
  \multicolumn{1}{r|}{0.57\%} \\ \cline{2-11} 
\multicolumn{1}{|c|}{} &
  \multicolumn{1}{c|}{10} &
  \multicolumn{1}{l|}{90.00\%} &
  \multicolumn{1}{r|}{93.10\%} &
  \multicolumn{1}{r|}{6.90\%} &
  \multicolumn{1}{l|}{61.33\%} &
  \multicolumn{1}{l|}{57.00\%} &
  \multicolumn{1}{l|}{43.00\%} &
  \multicolumn{1}{l|}{63.01\%} &
  \multicolumn{1}{r|}{0.54\%} &
  \multicolumn{1}{r|}{0.45\%} \\ \cline{2-11} 
\multicolumn{1}{|c|}{} &
  \multicolumn{1}{c|}{15} &
  \multicolumn{1}{l|}{95.10\%} &
  \multicolumn{1}{r|}{95.16\%} &
  \multicolumn{1}{r|}{4.84\%} &
  \multicolumn{1}{l|}{64.75\%} &
  \multicolumn{1}{l|}{39.00\%} &
  \multicolumn{1}{l|}{61.00\%} &
  \multicolumn{1}{l|}{64.44\%} &
  \multicolumn{1}{r|}{0.47\%} &
  \multicolumn{1}{r|}{0.52\%} \\ \cline{2-11} 
\multicolumn{1}{|c|}{} &
  \multicolumn{1}{c|}{20} &
  \multicolumn{1}{l|}{99.04\%} &
  \multicolumn{1}{r|}{99.48\%} &
  \multicolumn{1}{r|}{0.52\%} &
  \multicolumn{1}{l|}{66.44\%} &
  \multicolumn{1}{l|}{47.00\%} &
  \multicolumn{1}{l|}{53.00\%} &
  \multicolumn{1}{l|}{74.14\%} &
  \multicolumn{1}{r|}{0.67\%} &
  \multicolumn{1}{r|}{0.32\%} \\ \cline{2-11} 
\multicolumn{1}{|c|}{} &
  \multicolumn{1}{c|}{25} &
  \multicolumn{1}{l|}{99.26\%} &
  \multicolumn{1}{r|}{99.66\%} &
  \multicolumn{1}{r|}{0.34\%} &
  \multicolumn{1}{l|}{68.59\%} &
  \multicolumn{1}{l|}{73.00\%} &
  \multicolumn{1}{l|}{27.00\%} &
  \multicolumn{1}{l|}{82.72\%} &
  \multicolumn{1}{r|}{0.85\%} &
  \multicolumn{1}{r|}{0.14\%} \\ \hline
\multicolumn{1}{|c|}{\multirow{5}{*}{Flooding}} &
  \multicolumn{1}{c|}{5} &
  \multicolumn{1}{l|}{76.12\%} &
  \multicolumn{1}{r|}{63.44\%} &
  \multicolumn{1}{r|}{36.56\%} &
  \multicolumn{1}{l|}{58.43\%} &
  \multicolumn{1}{l|}{45.00\%} &
  \multicolumn{1}{l|}{55.00\%} &
  \multicolumn{1}{l|}{66.21\%} &
  \multicolumn{1}{r|}{0.59\%} &
  \multicolumn{1}{r|}{0.40\%} \\ \cline{2-11} 
\multicolumn{1}{|c|}{} &
  \multicolumn{1}{c|}{10} &
  \multicolumn{1}{l|}{97.78\%} &
  \multicolumn{1}{r|}{98.14\%} &
  \multicolumn{1}{r|}{1.86\%} &
  \multicolumn{1}{l|}{78.67\%} &
  \multicolumn{1}{l|}{62.00\%} &
  \multicolumn{1}{l|}{38.00\%} &
  \multicolumn{1}{l|}{86.76\%} &
  \multicolumn{1}{r|}{0.79\%} &
  \multicolumn{1}{r|}{0.20\%} \\ \cline{2-11} 
\multicolumn{1}{|c|}{} &
  \multicolumn{1}{c|}{15} &
  \multicolumn{1}{l|}{98.30\%} &
  \multicolumn{1}{r|}{98.79\%} &
  \multicolumn{1}{r|}{1.21\%} &
  \multicolumn{1}{l|}{80.22\%} &
  \multicolumn{1}{l|}{80.00\%} &
  \multicolumn{1}{l|}{20.00\%} &  
  \multicolumn{1}{l|}{86.84\%} &
  \multicolumn{1}{r|}{0.83\%} &
  \multicolumn{1}{r|}{0.16\%} \\ \cline{2-11} 
\multicolumn{1}{|c|}{} &
  \multicolumn{1}{c|}{20} &
  \multicolumn{1}{l|}{99.26\%} &
  \multicolumn{1}{r|}{99.48\%} &
  \multicolumn{1}{r|}{0.52\%} &
  \multicolumn{1}{l|}{80.30\%} &    
  \multicolumn{1}{l|}{66.00\%} &
  \multicolumn{1}{l|}{34.00\%} &
  \multicolumn{1}{l|}{87.89\%} &
  \multicolumn{1}{r|}{0.85\%} &
  \multicolumn{1}{r|}{0.14\%} \\ \cline{2-11} 
\multicolumn{1}{|c|}{} &
  \multicolumn{1}{c|}{25} &
  \multicolumn{1}{l|}{99.33\%} &
  \multicolumn{1}{r|}{99.61\%} &
  \multicolumn{1}{r|}{0.39\%} &
  \multicolumn{1}{l|}{74.74\%} &
  \multicolumn{1}{l|}{74.00\%} &
  \multicolumn{1}{l|}{26.00\%} &
  \multicolumn{1}{l|}{89.37\%} &
  \multicolumn{1}{r|}{0.86\%} &
  \multicolumn{1}{r|}{0.13\%} \\ \hline
\multicolumn{1}{l}{} &
  \multicolumn{1}{l}{} &
   &
  \multicolumn{1}{l}{} &
  \multicolumn{1}{l}{} &
   &
   &
   &
   &
  \multicolumn{1}{l}{} &
  \multicolumn{1}{l}{} \\
\multicolumn{1}{l}{} &
  \multicolumn{1}{l}{} &
   &
  \multicolumn{1}{l}{} &
  \multicolumn{1}{l}{} &
   &
   &
   &
   &
  \multicolumn{1}{l}{} &
  \multicolumn{1}{l}{} \\
\multicolumn{1}{l}{} &
  \multicolumn{1}{l}{} &
   &
  \multicolumn{1}{l}{} &
  \multicolumn{1}{l}{} &
   &
   &
   &
   &
  \multicolumn{1}{l}{} &
  \multicolumn{1}{l}{}

\end{tabular}}
\end{table}

In summary, FL-IDS shows high performance in detecting a range of network attacks, including black-hole, sinkhole, and flood attacks. In particular, when trained with the BTSC, FL-IDS even approaches the performance level of traditional Centralized IDS (C-IDS). This is achieved without the need to centrally collect all local data, showcasing the remarkable effectiveness of federated learning in network security. Through FL-IDS, local clients can learn and adapt independently while benefiting from global model updates. In addition to improving overall performance, this collaboration helps address complex and evolving threats that individual clients may not have encountered. With periodic retraining and an ongoing evolution of the global model, FL-IDS ensures that it stays current and effective in countering both existing and emerging threats, while seamlessly welcoming new nodes into the network.

\subsection{Communication and Response} 
While C-IDS demonstrates superior detection performance,  it achieves this by periodically collecting local data from each node to have a comprehensive view of the network. However, this comes with a communication cost. For C-IDS, the communication cost (CC) for the training process can be calculated using the formula:

\myrev{
\[CC = N * F * S * Periods\]}
where N is the number of UAVs (clients), F is the number of features collected per UAV, S is the size of each feature in bytes, and Periods is the number of data collection intervals.  In our experiments, the network consisted of 50 clients/nodes, each collecting 31 features, with each feature occupying 4 bytes. Data were collected over a total duration of 1800 seconds, consisting of 360 collection periods, each lasting 5 seconds, during the benign-only scenario. However, when attackers are included, the total number of collection periods doubles to 720, resulting in a total data collection duration of 3600 seconds.  \myrev{The communication cost is calculated as CC = 4,464,000  bytes.} This cost will increase proportionally with the duration of the operation or if additional features are required to address new types of attacks, such as lower layer features to detect malicious packet drop \cite{canbalaban2020cross}.

In contrast, FL-IDS only collects weights from each client to contribute the global model during (re-)training. \myrev{Although it requires every weight value in the model architecture periodically, the BTSC strategy (e.g., the top 20\% based on detection accuracy) further minimizes this cost by limiting training participation to a subset of the most effective nodes, thus reducing communication overhead. The communication cost can be expressed as:}

\myrev{\[CC_F = 2 * N * W * S * Epoch\]
}

\myrev{where N is the number of participating UAVs, W represents the weights of the transmitted model, S is the size of each weight in bytes and Epoch is the number of the training cycle. The factor of 2 accounts for both uploading the local model updates to the server and downloading the global model from the server. In our experiment, \(CC_F\) is calculated as 3,336,000 byte for BTSC.  }

\myrev{FL-IDS can result in higher costs when using complex models, primarily due to the transmission of large model weights or the requirement of a greater number of training epochs. However, FL-IDS eliminates the need for continuous data transmission in testing, as required by C-IDS, thereby enhancing efficiency for real-time detection scenarios. FL-IDS also offers inherent advantages for dynamic environments like FANETs. Given the high mobility and packet loss in such networks, FL’s ability to handle missing data during training ensures robust model updates without requiring constant retransmissions \cite{che2023multimodal}.} Furthermore, FL-IDS enables faster response times to security threats, as its decentralized structure eliminates the need for frequent data transmission to a central server. This allows real-time threat detection and mitigation without the latency associated with centralized data processing.

\subsection{Resource Consumption} 
Energy efficiency is a critical consideration in FL-IDS, particularly during the training process. UAVs perform local training using onboard datasets, ensuring data privacy by avoiding raw data transmission to a central server. While this approach enhances privacy, it can cause computational and energy costs, which depend on factors such as model complexity, dataset size, and computational capabilities of the UAVs. 

\myrev{To evaluate the energy overhead of the proposed scheme, we qualitatively estimate the per-node energy consumption as the sum of the energy required for local training and the communication energy for sharing model weights during the aggregation process. The energy consumption of communication is directly proportional to the communication cost, which can be represented as:
\[E_{com} = W * S * Epoch\]
The total energy consumption for each node, denoted as $E_{total}$, is calculated as the sum of the energy consumed during local training ($E_t$) and the communication energy ($E_{com}$): }

\myrev{\[E_{total} =  E_t + E_{com} \]}

To optimize energy consumption in the proposed scheme, strategies can be applied, such as increasing the interval between communication rounds, reducing the frequency of data collection, or employing energy-efficient machine learning techniques. These strategies aim to reduce energy usage while maintaining a balance between conserving energy and achieving high model accuracy. Careful tuning of these parameters is necessary to appropriately manage the trade-off between energy efficiency and model performance.

The BTSC strategy further reduces energy consumption by involving only the best-performing UAVs in training, conserving energy for non-participating nodes. Other promising techniques include split learning \cite{wang2022fedlite}, where only part of the model is trained locally while the rest resides on a central server, and few-shot learning \cite{song2023comprehensive}, which minimizes training data requirements.

\myrev{For comparison, L-IDS requires each node to independently train and maintain its model, with energy consumption represented by $E_t$. Unlike FL-IDS, L-IDS does not rely on data transmission to a central server, leading to a reduction in energy overhead. However, FL-IDS with the BTSC strategy distributes the workload more efficiently, reducing energy consumption across the network. In contrast, C-IDS incurs significantly higher energy costs because of the periodic transmission of large data and the centralized processing requirements at the GBS. In our experiment, the energy overhead of C-IDS is calculated based on $CC * \epsilon$. This centralized approach, particularly in wireless UAV networks, contributes substantially to energy expenditure \cite{brik2020federated}. Furthermore, C-IDS faces storage challenges due to the need to store large volumes of data collected periodically.}

\subsection{Privacy and Security} While Central IDS (C-IDS) generally exhibits high performance in detecting attacks, it necessitates the central collection of local data, which raises significant concerns about data privacy and security. This data collected could contain sensitive information, such as communication content,  operation-specific or node-specific information, and location data. In addition, a single C-IDS system can become an attractive target for potential attackers. If the system is compromised, attackers could gain access to all this sensitive data.

In contrast, Federated IDS (FL-IDS) offers a decentralized and privacy-focused approach. With FL-IDS, each node leverages its local data to train individual local models and contributes to the global model by sharing only its local model's weights. As a result, FL-IDS provides a robust and privacy-conscious solution that not only minimizes data privacy and security concerns but also leverages the collective intelligence of local nodes for effective intrusion detection.

Although FL has a lot of potential benefits and provides privacy, it could still be susceptible to adversarial attacks, much like other types of ML methods. Adversaries may attempt to undermine the integrity of the FL process by providing erroneous or misleading data. This could involve intentionally providing inaccurate or deceptive data with the goal of misleading the trained model. Additionally, adversaries might engage in the malicious engineering of data, where they intentionally manipulate data in a deceptive manner to trick the trained model into making mistakes \cite{liu2022threats}.

\myrev{To ensure secure model transmission over in-band communication in federated learning-based IDS, several security mechanisms can be employed.  First, end-to-end encryption can be implemented to maintain the confidentiality and integrity of models during transmission, preventing adversaries from intercepting or tampering with them \cite{li2019end}. Second, the adoption of blockchain technology can establish a tamper-proof record of model updates, enhancing transparency and allowing nodes to track the history of model changes \cite{9420262}. 
Third, robust anomaly detection mechanisms can be deployed to validate received model updates, identifying and discarding malicious or suspicious contributions before they affect the global model \cite{bouacida2021vulnerabilities}. These mechanisms analyze update distributions to detect deviations indicative of poisoning or model manipulation, thus ensuring the integrity of the global model. 
Additionally, techniques such as secure multiparty computation (SMC) and differential privacy (DP) can further enhance security. SMC allows multiple parties to securely aggregate updates without revealing individual inputs, preserving confidentiality throughout the aggregation process. DP, on the other hand, improves privacy by introducing noise to updates, preventing the identification of individual contributions, and mitigating potential privacy breaches while maintaining model performance.}

Although adversarial attacks are a common issue for all types of IDS; C-IDS and F-IDS also introduce a single point of failure due to collecting and/or processing data at the GBS. On the other hand, while the global model can be moved to another node, C-IDS requires the existence of a powerful node due to processing and storing large amount of data. In the case of L-IDS, where each node operates independently, the decentralized approach limits the potential impact of a compromise to individual nodes. However, the lack of inter-node communication may also restrict the system's ability to implement collective defense mechanisms or coordinated responses to sophisticated attacks, potentially limiting its overall resilience against certain types of threat.

\section{Conclusion}
\label{sec:conclusion}
This study introduces a novel approach to intrusion detection in FANETs by proposing Federated IDS (FL-IDS). Given the distributed nature of network data in FANETs, a distributed and cooperative architecture is inherently more suitable. However, in such an architecture, data sharing raises privacy concerns, particularly in the context of UAVs engaged in mission-critical applications. FL-IDS offers a solution to address these concerns and, in this study, it is introduced and compared comprehensively with traditional approaches. C-IDS and L-IDS.

The comparative evaluation covers multiple metrics, including effectiveness, communication and response, resource consumption, privacy and security considerations. The assessment extends to various networks with varying mobility patterns to ensure a comprehensive analysis. In addition, our study incorporates realistic network scenarios that account for factors such as 3D node movement, local data collection by each node, and realistic traffic patterns. To our knowledge, this study represents one of the few recent studies in the field of FANET security \cite{chulerttiyawong2023sybil}\cite{zhai2023etd} where data are based on an actual FANET dataset, but with more realistic settings specifically tailored to suit the dynamics and requirements of FANETs. In previous studies \cite{ferrag2021federated}, the application of federated learning was often demonstrated using synthetic data that mimicked federations of IoT data. We believe that this study contributes significantly to the field as it represents the first comprehensive exploration of the use of federated learning for routing attacks and offers a detailed comparison with traditional intrusion detection approaches. The experimental results demonstrate the effectiveness and suitability of the proposed FL-IDS approach driven by its distributed training methodology, while also addressing critical privacy concerns.

%% Use \subsubsection, \paragraph, \subparagraph commands to 
%% start 3rd, 4th and 5th level sections.
%% Refer following link for more details.
%% https://en.wikibooks.org/wiki/LaTeX/Document_Structure#Sectioning_commands

\bibliographystyle{elsarticle-num} 
\bibliography{references}

\begin{thebibliography}{10}
\expandafter\ifx\csname url\endcsname\relax
  \def\url#1{\texttt{#1}}\fi
\expandafter\ifx\csname urlprefix\endcsname\relax\def\urlprefix{URL }\fi
\expandafter\ifx\csname href\endcsname\relax
  \def\href#1#2{#2} \def\path#1{#1}\fi

\bibitem{bekmezci2013flying}
I.~Bekmezci, O.~K. Sahingoz, {\c{S}}.~Temel, Flying ad-hoc networks (fanets): A survey, Ad Hoc Networks 11~(3) (2013) 1254--1270.

\bibitem{boursianis2022internet}
A.~D. Boursianis, M.~S. Papadopoulou, P.~Diamantoulakis, A.~Liopa-Tsakalidi, P.~Barouchas, G.~Salahas, G.~Karagiannidis, S.~Wan, S.~K. Goudos, Internet of things (iot) and agricultural unmanned aerial vehicles (uavs) in smart farming: A comprehensive review, Internet of Things 18 (2022) 100187.

\bibitem{kakamoukas2022fanets}
G.~A. Kakamoukas, P.~G. Sarigiannidis, A.~A. Economides, Fanets in agriculture-a routing protocol survey, Internet of Things 18 (2022) 100183.

\bibitem{yaacoub2020security}
J.-P. Yaacoub, H.~Noura, O.~Salman, A.~Chehab, Security analysis of drones systems: Attacks, limitations, and recommendations, Internet of Things 11 (2020) 100218.

\bibitem{sharma2023secure}
J.~Sharma, P.~S. Mehra, Secure communication in iot-based uav networks: A systematic survey, Internet of Things (2023) 100883.

\bibitem{tsao2022survey}
K.-Y. Tsao, T.~Girdler, V.~G. Vassilakis, A survey of cyber security threats and solutions for uav communications and flying ad-hoc networks, Ad Hoc Networks 133 (2022) 102894.

\bibitem{chriki2019fanet}
A.~Chriki, H.~Touati, H.~Snoussi, F.~Kamoun, Fanet: Communication, mobility models and security issues, Computer Networks 163 (2019) 106877.

\bibitem{amponis2021survey}
G.~Amponis, T.~Lagkas, P.~Sarigiannidis, V.~Vitsas, P.~Fouliras, S.~Wan, A survey on fanet routing from a cross-layer design perspective, Journal of Systems Architecture 120 (2021) 102281.

\bibitem{ceviz2021analysis}
O.~Ceviz, P.~Sadioglu, S.~Sen, Analysis of routing attacks in fanets, in: International Conference on Ad Hoc Networks, Springer, 2021, pp. 3--17.

\bibitem{nayfeh2023machine}
M.~Nayfeh, Y.~Li, K.~Al~Shamaileh, V.~Devabhaktuni, N.~Kaabouch, Machine learning modeling of gps features with applications to uav location spoofing detection and classification, Computers \& Security 126 (2023) 103085.

\bibitem{Ouiazzane2020}
S.~Ouiazzane, F.~Barramou, M.~Addou, {Towards a Multi-Agent based Network Intrusion Detection System for a Fleet of Drones}, Int. J. Adv. Comput. Sci. Appl. 11~(10) (2020) 351--362.
\newblock \href {https://doi.org/10.14569/IJACSA.2020.0111044} {\path{doi:10.14569/IJACSA.2020.0111044}}.

\bibitem{zhang2021survey}
C.~Zhang, Y.~Xie, H.~Bai, B.~Yu, W.~Li, Y.~Gao, A survey on federated learning, Knowledge-Based Systems 216 (2021) 106775.

\bibitem{baig2022securing}
Z.~Baig, N.~Syed, N.~Mohammad, Securing the smart city airspace: Drone cyber attack detection through machine learning, Future Internet 14~(7) (2022) 205.

\bibitem{Dat}
Labs, \href{https://www.vtolabs.com/drone-forensics (accessed on 15 April 2022)}{V. drone forensics} (2020).
\newline\urlprefix\url{https://www.vtolabs.com/drone-forensics (accessed on 15 April 2022)}

\bibitem{da2022development}
L.~M. da~Silva, H.~B. d.~B. Menezes, M.~d.~S. Luccas, C.~Mailer, A.~S.~R. Pinto, A.~Boava, M.~Rodrigues, I.~G. Ferr{\~a}o, J.~C. Estrella, K.~R. L. J.~C. Branco, Development of an efficiency platform based on mqtt for uav controlling and dos attack detection, Sensors 22~(17) (2022) 6567.

\bibitem{kolias2015intrusion}
C.~Kolias, G.~Kambourakis, A.~Stavrou, S.~Gritzalis, Intrusion detection in 802.11 networks: Empirical evaluation of threats and a public dataset, IEEE Communications Surveys \& Tutorials 18~(1) (2015) 184--208.

\bibitem{shrestha2021machine}
R.~Shrestha, A.~Omidkar, S.~A. Roudi, R.~Abbas, S.~Kim, Machine-learning-enabled intrusion detection system for cellular connected uav networks, Electronics 10~(13) (2021) 1549.

\bibitem{chulerttiyawong2023sybil}
D.~Chulerttiyawong, A.~Jamalipour, Sybil attack detection in internet of flying things-ioft: A machine learning approach, IEEE Internet of Things Journal (2023).

\bibitem{varga2010omnet}
A.~Varga, Omnet++, in: Modeling and tools for network simulation, Springer, 2010, pp. 35--59.

\bibitem{zhai2023etd}
W.~Zhai, L.~Liu, Y.~Ding, S.~Sun, Y.~Gu, Etd: An efficient time delay attack detection framework for uav networks, IEEE Transactions on Information Forensics and Security (2023).

\bibitem{keranen2009one}
A.~Ker{\"a}nen, J.~Ott, T.~K{\"a}rkk{\"a}inen, The one simulator for dtn protocol evaluation, in: Proceedings of the 2nd international conference on simulation tools and techniques, 2009, pp. 1--10.

\bibitem{abu2022high}
Q.~Abu Al-Haija, A.~Al~Badawi, High-performance intrusion detection system for networked uavs via deep learning, Neural Computing and Applications 34~(13) (2022) 10885--10900.

\bibitem{zhao2018prediction}
L.~Zhao, A.~Alipour-Fanid, M.~Slawski, K.~Zeng, Prediction-time efficient classification using feature computational dependencies, in: Proceedings of the 24th ACM SIGKDD International Conference on Knowledge Discovery \& Data Mining, 2018, pp. 2787--2796.

\bibitem{Ramadan2021}
R.~A. Ramadan, A.~H. Emara, M.~Al-Sarem, M.~Elhamahmy, {Internet of drones intrusion detection using deep learning}, Electron. 10~(21) (2021).
\newblock \href {https://doi.org/10.3390/electronics10212633} {\path{doi:10.3390/electronics10212633}}.

\bibitem{kddcup1999}
{KDD Cup 1999}, Kdd cup 1999, \url{http://kdd.ics.uci.edu/databases/kddcup99/kddcup99.html} (October 2007).

\bibitem{5356528}
M.~Tavallaee, E.~Bagheri, W.~Lu, A.~A. Ghorbani, A detailed analysis of the kdd cup 99 data set, in: 2009 IEEE Symposium on Computational Intelligence for Security and Defense Applications, 2009, pp. 1--6.
\newblock \href {https://doi.org/10.1109/CISDA.2009.5356528} {\path{doi:10.1109/CISDA.2009.5356528}}.

\bibitem{bouhamed2021lightweight}
O.~Bouhamed, O.~Bouachir, M.~Aloqaily, I.~Al~Ridhawi, Lightweight ids for uav networks: A periodic deep reinforcement learning-based approach, in: 2021 IFIP/IEEE International Symposium on Integrated Network Management (IM), IEEE, 2021, pp. 1032--1037.

\bibitem{panigrahi2018detailed}
R.~Panigrahi, S.~Borah, A detailed analysis of cicids2017 dataset for designing intrusion detection systems, International Journal of Engineering \& Technology 7~(3.24) (2018) 479--482.

\bibitem{arthur2019detecting}
M.~P. Arthur, Detecting signal spoofing and jamming attacks in uav networks using a lightweight ids, in: 2019 international conference on computer, information and telecommunication systems (CITS), IEEE, 2019, pp. 1--5.

\bibitem{mowla2019federated}
N.~I. Mowla, N.~H. Tran, I.~Doh, K.~Chae, Federated learning-based cognitive detection of jamming attack in flying ad-hoc network, IEEE Access 8 (2019) 4338--4350.

\bibitem{mowla2020afrl}
N.~I. Mowla, N.~H. Tran, I.~Doh, K.~Chae, Afrl: Adaptive federated reinforcement learning for intelligent jamming defense in fanet, Journal of Communications and Networks 22~(3) (2020) 244--258.

\bibitem{da2023anomaly}
L.~M. Da~Silva, I.~G. Ferr{\~a}o, C.~Dezan, D.~Espes, K.~R. Branco, Anomaly-based intrusion detection system for in-flight and network security in uav swarm, in: 2023 International Conference on Unmanned Aircraft Systems (ICUAS), IEEE, 2023, pp. 812--819.

\bibitem{Ns-3}
The ns-3 network simulator, \url{http://www.nsnam.org/}.

\bibitem{punal2014crawdad}
O.~Pu{\~n}al, C.~Pereira, A.~Aguiar, J.~Gross, Crawdad dataset uportorwthaachen/vanetjamming2014 (v. 2014-05-12) (2014).

\bibitem{whelan2020novelty}
J.~Whelan, T.~Sangarapillai, O.~Minawi, A.~Almehmadi, K.~El-Khatib, Novelty-based intrusion detection of sensor attacks on unmanned aerial vehicles, in: Proceedings of the 16th ACM symposium on QoS and security for wireless and mobile networks, 2020, pp. 23--28.

\bibitem{mcmahan2017communication}
B.~McMahan, E.~Moore, D.~Ramage, S.~Hampson, B.~A. y~Arcas, Communication-efficient learning of deep networks from decentralized data, in: Artificial intelligence and statistics, PMLR, 2017, pp. 1273--1282.

\bibitem{reddi2020adaptive}
S.~Reddi, Z.~Charles, M.~Zaheer, Z.~Garrett, K.~Rush, J.~Kone{\v{c}}n{\`y}, S.~Kumar, H.~B. McMahan, Adaptive federated optimization, arXiv preprint arXiv:2003.00295 (2020).

\bibitem{CAMPOS2022108661}
E.~M. Campos, P.~F. Saura, A.~González-Vidal, J.~L. Hernández-Ramos, J.~B. Bernabé, G.~Baldini, A.~Skarmeta, \href{https://www.sciencedirect.com/science/article/pii/S1389128621005405}{Evaluating federated learning for intrusion detection in internet of things: Review and challenges}, Computer Networks 203 (2022) 108661.
\newblock \href {https://doi.org/https://doi.org/10.1016/j.comnet.2021.108661} {\path{doi:https://doi.org/10.1016/j.comnet.2021.108661}}.
\newline\urlprefix\url{https://www.sciencedirect.com/science/article/pii/S1389128621005405}

\bibitem{moustafa2019new}
N.~Moustafa, New generations of internet of things datasets for cybersecurity applications based machine learning: Ton\_iot datasets, in: Proceedings of the eResearch Australasia Conference, Brisbane, Australia, 2019, pp. 21--25.

\bibitem{yu2020fed+}
P.~Yu, L.~Wynter, S.~H. Lim, Fed+: A family of fusion algorithms for federated learning, arXiv preprint arXiv:2009.06303 (2020).

\bibitem{abou2023secure}
Z.~Abou El~Houda, H.~Moudoud, L.~Khoukhi, Secure and efficient federated learning for robust intrusion detection in iot networks, in: GLOBECOM 2023-2023 IEEE Global Communications Conference, IEEE, 2023, pp. 2668--2673.

\bibitem{abou2024blockchain}
Z.~Abou El~Houda, H.~Moudoud, B.~Brik, L.~Khoukhi, Blockchain-enabled federated learning for enhanced collaborative intrusion detection in vehicular edge computing, IEEE Transactions on Intelligent Transportation Systems (2024).

\bibitem{ferrag2021federated}
M.~A. Ferrag, O.~Friha, L.~Maglaras, H.~Janicke, L.~Shu, Federated deep learning for cyber security in the internet of things: Concepts, applications, and experimental analysis, IEEE Access 9 (2021) 138509--138542.

\bibitem{rahman2020internet}
S.~A. Rahman, H.~Tout, C.~Talhi, A.~Mourad, Internet of things intrusion detection: Centralized, on-device, or federated learning?, IEEE Network 34~(6) (2020) 310--317.

\bibitem{ALZAHRANI2020102706}
B.~Alzahrani, O.~S. Oubbati, A.~Barnawi, M.~Atiquzzaman, D.~Alghazzawi, \href{https://www.sciencedirect.com/science/article/pii/S1084804520301806}{Uav assistance paradigm: State-of-the-art in applications and challenges}, Journal of Network and Computer Applications 166 (2020) 102706.
\newblock \href {https://doi.org/https://doi.org/10.1016/j.jnca.2020.102706} {\path{doi:https://doi.org/10.1016/j.jnca.2020.102706}}.
\newline\urlprefix\url{https://www.sciencedirect.com/science/article/pii/S1084804520301806}

\bibitem{9275621}
F.~Jiang, K.~Wang, L.~Dong, C.~Pan, W.~Xu, K.~Yang, Ai driven heterogeneous mec system with uav assistance for dynamic environment: Challenges and solutions, IEEE Network 35~(1) (2021) 400--408.
\newblock \href {https://doi.org/10.1109/MNET.011.2000440} {\path{doi:10.1109/MNET.011.2000440}}.

\bibitem{sharafaldin2018toward}
I.~Sharafaldin, A.~H. Lashkari, A.~A. Ghorbani, Toward generating a new intrusion detection dataset and intrusion traffic characterization., ICISSp 1 (2018) 108--116.

\bibitem{moustafa2015unsw}
N.~Moustafa, J.~Slay, Unsw-nb15: a comprehensive data set for network intrusion detection systems (unsw-nb15 network data set), in: 2015 military communications and information systems conference (MilCIS), IEEE, 2015, pp. 1--6.

\bibitem{almomani2016wsn}
I.~Almomani, B.~Al-Kasasbeh, M.~Al-Akhras, et~al., Wsn-ds: A dataset for intrusion detection systems in wireless sensor networks, Journal of Sensors 2016 (2016).

\bibitem{koroniotis2019towards}
N.~Koroniotis, N.~Moustafa, E.~Sitnikova, B.~Turnbull, Towards the development of realistic botnet dataset in the internet of things for network forensic analytics: Bot-iot dataset, Future Generation Computer Systems 100 (2019) 779--796.

\bibitem{vaccari2020mqttset}
I.~Vaccari, G.~Chiola, M.~Aiello, M.~Mongelli, E.~Cambiaso, Mqttset, a new dataset for machine learning techniques on mqtt, Sensors 20~(22) (2020) 6578.

\bibitem{agrawal2022federated}
S.~Agrawal, S.~Sarkar, O.~Aouedi, G.~Yenduri, K.~Piamrat, M.~Alazab, S.~Bhattacharya, P.~K.~R. Maddikunta, T.~R. Gadekallu, Federated learning for intrusion detection system: Concepts, challenges and future directions, Computer Communications (2022).

\bibitem{Ullah2017}
H.~Ullah, M.~Abu-Tair, S.~McClean, P.~Nixon, G.~Parr, C.~Luo, {An unmanned aerial vehicle based wireless network for bridging communication}, in: Proc. - 14th Int. Symp. Pervasive Syst. Algorithms Networks, I-SPAN 2017, 11th Int. Conf. Front. Comput. Sci. Technol. FCST 2017 3rd Int. Symp. Creat. Comput. ISCC 2017, 2017.
\newblock \href {https://doi.org/10.1109/ISPAN-FCST-ISCC.2017.65} {\path{doi:10.1109/ISPAN-FCST-ISCC.2017.65}}.

\bibitem{che2023multimodal}
L.~Che, J.~Wang, Y.~Zhou, F.~Ma, Multimodal federated learning: A survey, Sensors 23~(15) (2023) 6986.

\bibitem{Daniel2014}
A.~Daniel, {A Survey on Detection of Sinkhole Attack in Wireless Sensor Networks} 91~(7) (2014) 48--52.

\bibitem{zhi2020security}
Y.~Zhi, Z.~Fu, X.~Sun, J.~Yu, Security and privacy issues of uav: a survey, Mobile Networks and Applications 25 (2020) 95--101.

\bibitem{bai2023towards}
Y.~Bai, H.~Zhao, X.~Zhang, Z.~Chang, R.~J{\"a}ntti, K.~Yang, Towards autonomous multi-uav wireless network: A survey of reinforcement learning-based approaches, IEEE Communications Surveys \& Tutorials (2023).

\bibitem{cui2021channel}
Z.~Cui, K.~Guan, C.~Briso-Rodríguez, B.~Ai, Z.~Zhong, C.~Oestges, Channel modeling for uav communications: State of the art, case studies, and future directions (2021).
\newblock \href {http://arxiv.org/abs/2012.06707} {\path{arXiv:2012.06707}}.

\bibitem{6127781}
J.~P. Rohrer, E.~K. Çetinkaya, H.~Narra, D.~Broyles, K.~Peters, J.~P. Sterbenz, Aerorp performance in highly-dynamic airborne networks using 3d gauss-markov mobility model, in: 2011 - MILCOM 2011 Military Communications Conference, 2011, pp. 834--841.
\newblock \href {https://doi.org/10.1109/MILCOM.2011.6127781} {\path{doi:10.1109/MILCOM.2011.6127781}}.

\bibitem{ceviz2023survey}
O.~Ceviz, P.~Sadioglu, S.~Sen, A survey of security in uavs and fanets: Issues, threats, analysis of attacks, and solutions, arXiv preprint arXiv:2306.14281 (2023).

\bibitem{sen2011evolutionary}
S.~Sen, J.~A. Clark, Evolutionary computation techniques for intrusion detection in mobile ad hoc networks, Computer Networks 55~(15) (2011) 3441--3457.

\bibitem{li2020federated}
T.~Li, A.~K. Sahu, M.~Zaheer, M.~Sanjabi, A.~Talwalkar, V.~Smith, Federated optimization in heterogeneous networks, Proceedings of Machine learning and systems 2 (2020) 429--450.

\bibitem{li2020preserving}
Z.~Li, V.~Sharma, S.~P. Mohanty, Preserving data privacy via federated learning: Challenges and solutions, IEEE Consumer Electronics Magazine 9~(3) (2020) 8--16.

\bibitem{canbalaban2020cross}
E.~Canbalaban, S.~Sen, A cross-layer intrusion detection system for rpl-based internet of things, in: Ad-Hoc, Mobile, and Wireless Networks: 19th International Conference on Ad-Hoc Networks and Wireless, ADHOC-NOW 2020, Bari, Italy, October 19--21, 2020, Proceedings 19, Springer, 2020, pp. 214--227.

\bibitem{wang2022fedlite}
J.~Wang, H.~Qi, A.~S. Rawat, S.~Reddi, S.~Waghmare, F.~X. Yu, G.~Joshi, Fedlite: A scalable approach for federated learning on resource-constrained clients, arXiv preprint arXiv:2201.11865 (2022).

\bibitem{song2023comprehensive}
Y.~Song, T.~Wang, P.~Cai, S.~K. Mondal, J.~P. Sahoo, A comprehensive survey of few-shot learning: Evolution, applications, challenges, and opportunities, ACM Computing Surveys (2023).

\bibitem{brik2020federated}
B.~Brik, A.~Ksentini, M.~Bouaziz, Federated learning for uavs-enabled wireless networks: Use cases, challenges, and open problems, IEEE Access 8 (2020) 53841--53849.

\bibitem{liu2022threats}
P.~Liu, X.~Xu, W.~Wang, Threats, attacks and defenses to federated learning: issues, taxonomy and perspectives, Cybersecurity 5~(1) (2022) 1--19.

\bibitem{li2019end}
H.~Li, T.~Han, An end-to-end encrypted neural network for gradient updates transmission in federated learning, arXiv preprint arXiv:1908.08340 (2019).

\bibitem{9420262}
H.~Liu, S.~Zhang, P.~Zhang, X.~Zhou, X.~Shao, G.~Pu, Y.~Zhang, Blockchain and federated learning for collaborative intrusion detection in vehicular edge computing, IEEE Transactions on Vehicular Technology 70~(6) (2021) 6073--6084.
\newblock \href {https://doi.org/10.1109/TVT.2021.3076780} {\path{doi:10.1109/TVT.2021.3076780}}.

\bibitem{bouacida2021vulnerabilities}
N.~Bouacida, P.~Mohapatra, Vulnerabilities in federated learning, IEEE Access 9 (2021) 63229--63249.

\end{thebibliography}

\end{document}